\documentclass[conference,compsoc]{IEEEtran}
\IEEEoverridecommandlockouts

\ifCLASSOPTIONcompsoc
  \usepackage[nocompress]{cite}
\else
  \usepackage{cite}
\fi

\usepackage{amsmath,amssymb,amsfonts}
\usepackage{algorithmic}
\usepackage{graphicx}
\usepackage{textcomp}
\usepackage{xcolor}
\usepackage{multirow}
\usepackage{array}
\usepackage{booktabs} 
\usepackage{makecell} 
\usepackage{tabularx}
\usepackage{paralist}
\usepackage[dvipsnames]{xcolor}
\usepackage[most]{tcolorbox}
\usepackage{enumitem}
\usepackage{pifont}
\usepackage{adjustbox}
\usepackage{wasysym}
\usepackage{orcidlink}
\usepackage{marvosym}
\usepackage{ifthen}
\usepackage{xurl} 
\usepackage{float}
\usepackage{rotating}
\usepackage{colortbl}
\usepackage{diagbox}
\usepackage{svg}
\usepackage{moresize}

\newcolumntype{Y}{>{\raggedright\arraybackslash}X}
\newcolumntype{C}[1]{>{\centering\arraybackslash}p{#1}}

\definecolor{ieeeblue}{RGB}{26, 12, 171}
\definecolor{sectionColor}{HTML}{B20100}
\hypersetup{
    colorlinks=true,
    linkcolor=sectionColor,
    citecolor=ieeeblue,
    filecolor=ieeeblue,
    urlcolor=black
}

\newcommand*\nogap[1]{}    

\newcommand{\cmark}{\ding{51}} 
\newcommand{\xmark}{\ding{55}} 

\newtcbox{\trend}{%
  on line,              
  boxsep=1.7pt,           
  left=1pt,             
  right=1pt,            
  top=1pt,              
  bottom=1pt,           
  colback=Cerulean!20,      
  arc=6pt,              
  boxrule=0pt,        
}

\newtcbox{\limitation}{%
  on line,              
  boxsep=1.7pt,           
  left=1pt,             
  right=1pt,            
  top=1pt,              
  bottom=1pt,           
  colback=orange!20,      
  arc=6pt,              
  boxrule=0pt,        
}

\newtcbox{\issue}{%
  on line,              
  boxsep=1.7pt,           
  left=1pt,             
  right=1pt,            
  top=1pt,              
  bottom=1pt,           
  colback=red!20,      
  arc=6pt,              
  boxrule=0pt,        
}

\newtcbox{\recommendation}{%
  on line,              
  boxsep=1.7pt,           
  left=1pt,             
  right=1pt,            
  top=1pt,              
  bottom=1pt,           
  colback=Purple!20,      
  arc=6pt,              
  boxrule=0pt,        
}

\newtcbox{\opportunity}{%
  on line,              
  boxsep=1.7pt,           
  left=1pt,             
  right=1pt,            
  top=1pt,              
  bottom=1pt,           
  colback=SpringGreen!40,      
  arc=6pt,              
  boxrule=0pt,        
}

\def\BibTeX{{\rm B\kern-.05em{\sc i\kern-.025em b}\kern-.08em
    T\kern-.1667em\lower.7ex\hbox{E}\kern-.125emX}}
\begin{document}

\title{{\fontsize{11}{12}\selectfont \textnormal{To appear in the IEEE Symposium on Security \& Privacy, May 2026}} \\[1ex]  SoK: After Decades of Web Tracker Detection, What's Next?
\vspace{-1.5em}
}

\newcommand\copyrighttext{
  \footnotesize \textcopyright \kern1pt 2026 IEEE. Personal use of this material is permitted. Permission from IEEE must be obtained for all other uses, in any current or future media, including reprinting/republishing this material for advertising or promotional purposes, creating new collective works, for resale or redistribution to servers or lists, or reuse of any copyrighted component of this work in other works. 
}
\newcommand\copyrightnotice{
  \begin{tikzpicture}[remember picture,overlay]
    \node[anchor=south,yshift=10pt] at (current page.south)
    {\fbox{\parbox{\dimexpr\textwidth-\fboxsep-\fboxrule\relax}{\copyrighttext}}};
  \end{tikzpicture}
}

\author{\IEEEauthorblockN{Wolf Rieder\kern1pt\orcidlink{0009-0001-4932-9814}\kern0.7pt\textsuperscript{\href{mailto:w.rieder@tu-berlin.de}{\Letter}}$^{\dagger}$,
Philip Raschke\kern1pt\orcidlink{0000-0002-6738-7137}$^{\dagger}$, 
Thomas Cory\kern1pt\orcidlink{0000-0002-3452-9944}$^{\dagger}$, \\
Christian René Sechting\kern1pt\orcidlink{0009-0008-8869-3396}$^{\dagger}$, 
Aditya Kumar\kern1pt\orcidlink{0000-0001-6590-8704}$^{\parallel}$ and 
Axel Küpper\kern1pt\orcidlink{0000-0002-4356-5613}$^{\dagger}$}
\IEEEauthorblockA{$^{\dagger}$Technische Universität Berlin $^{\parallel}$Independent Researcher}
}

\maketitle
\copyrightnotice


\vspace{-1.25em}
\begin{abstract}
Web tracking is an omnipresent phenomenon in today's web, affecting users in their day-to-day lives. Filter lists and blockers were invented to detect trackers and to protect users. Due to limitations of said tools, researchers developed web tracker detectors to replace them. 
No review constructed a universal perspective and classification of web tracker detectors until now. Past reviews focused either on the field as a whole or on web tracking techniques.  
In this SoK paper, we present the most comprehensive meta-science study on web tracker detection by systematizing and synthesizing the available knowledge. We conduct a systematic review, resulting in 59 primary and 16 supplementary studies out of a corpus of 832 papers. Based on these findings we suggest a taxonomy, observe and evaluate trends, propose open research gaps, and recommendations with which we aim to lay the foundations for future web tracker detection research. In addition, we conduct a limited reproducibility study to assess the validity of past studies and highlight emerging problems in this field. 
\end{abstract}


\section{Introduction}\label{sec:intro}
Web tracker detection (detection) approaches have been proposed as a potential replacement for prominent filter lists such as \textsc{EasyList}~\cite{easyList} and \textsc{EasyPrivacy}~\cite{easyPrivacy} for live detection or to automatically generate these lists. This eliminates the need for community members to manually build them, which can lead to known drawbacks~\cite{alrizah, hashmi, snyder}. Existing research is primarily focused on the development or identification of new web tracking techniques, e.g. identification of user; and technologies (techniques and technologies), such as fingerprinting~\cite{laperdrix, fpguard, iqbal-fingerprinting-2021}; or new web tracker detectors (detector)~\cite{amjad-2024, lee_adflush_2024, rieder_beyond_2025}. Given the observation of the first web tracker in 1996~\cite{lerner} as well as the maturity and active research in this field, a plethora of detectors with their respective datasets, metrics, and features have been developed over the past three decades. However, there has been limited research on a holistic and historical picture of the development and future of detectors. 

Consequently, a clear and systematic picture on the intricacies, challenges, problems, and hidden insights in detection research remains missing. The lack of such systematization in an established, major research area is a significant gap in the literature. 
This Systematization of Knowledge (SoK) paper aims to address this gap and to provide insights by applying a meta-science perspective. 

Prior to starting our SoK paper, we searched through Google Scholar~\cite{google_scholar} in October 2024 using search terms related to detection and (systematic) reviews without language restrictions. There are reviews and surveys on web tracking and its intricacies such as techniques and technologies published before 2025~\cite{Bujlow}. However, not one review aimed to systematize all or nearly all available detector studies in a methodical way (see~\autoref{sec:rel-work}). The minority of studies reviewed detectors specifically, and only then selected an arbitrary number of papers, leaving the majority of published results out -- a sampling bias which has not been addressed. 

Lastly, we want to point out the overall effectiveness of past detectors as our final motivation. We found that both simple and complex detectors have consistently achieved performance metrics in the 80th or even 90th percentile, showing that the detection of trackers has been working well for years. This overall trend begs the question: How will this field evolve in the future, if the detection itself is seemingly not a problem anymore? For that, we have to first understand the past and the present state of web tracker detection to infer its potential future path. 

\begin{table}[ht]
    \centering
    \caption{SoK Paper Scope based on the PICOC Framework.}
    \scriptsize
	\begin{tabularx}{\columnwidth}{rX}
            \toprule
		\textbf{Criteria} & \textbf{Description} 
            \\
		\midrule
            \textbf{Population} 
            & Empirical research studies on detection in the context of web privacy and web tracking. 
            \vspace{2pt}
            \\ 
            \textbf{Intervention} 
            & Detection approaches that develop and evaluate a detector, based on heuristics or learned models, excluding URL-based filtering approaches. 
            \\    
            \textbf{Comparisons} 
            & Different detector designs and study methodologies across a typical machine learning-workflow. 
            \\ 
            \textbf{Outcome} 
            & A systematization of detector and study characteristics, and a meta-scientific appraisal of the reported evidence. 
            \\ 
            \textbf{Context} 
            & Web privacy researchers from academia and industry that develop detectors or privacy defenses. 
            \\ 
		\bottomrule
	\end{tabularx}
 \label{table:picoc}
\end{table}

\textbf{Scope.}
Our objectives are: to systematically analyze and categorize detection research by characterizing study and detector properties, to develop recommendations, and to derive a future research agenda, all in a transparent and reproducible manner. We tackle this objective and the aforementioned gaps in the literature, by first determining the scope and boundaries of our SoK as shown in Table~\ref{table:picoc} with the recommended Population, Intervention, Comparison, Outcome, and Context (PICOC) framework~\cite{kitchenham}. 

Based on the outlined scope, our SoK paper aims to answer the following research questions (RQ): 

\begin{enumerate}[itemsep=0.3mm, parsep=0pt]
    \item[\textit{RQ1}] What study and detector characteristics define web tracker detection research, and how are detectors designed in the literature? (\autoref{sec:prelim}, \autoref{sec:taxonomy-final}, \autoref{sec:sys}) 
    \item[\textit{RQ2}] Which similarities, differences, and historical trends can be observed across web tracker detection studies over time? (\autoref{sec:taxonomy-final}, \autoref{sec:sys}) 
    \item[\textit{RQ3}] How valid, reliable, and reproducible are performance evaluations in web tracker detection research, and how completely are they reported? (\autoref{sec:performance-evaluation}, \autoref{sec:rep}, Appendix~\ref{sec:quality-results})
    \item[\textit{RQ4}] What are the existing knowledge gaps, future research directions, and recommendations for web tracker detection research? (\autoref{sec:sys}, \autoref{sec:rep}, \autoref{sec:discussion})
\end{enumerate}

To answer our RQs, we contribute the following: 

\begin{itemize}[itemsep=0.3mm, parsep=0pt]
    \item \textbf{Systematization of knowledge.} We present and systematize detection research from different perspectives such as stakeholders and classification pipeline. Our review included 59 primary and 16 supplementary studies from the last three decades, which is the largest review on this topic to date (\autoref{sec:prelim}, \autoref{sec:sys}, \autoref{sec:discussion}). 
    \item \textbf{Taxonomy.} The first comprehensive taxonomy that classifies detectors across nine categories, 29 dimensions, and 96 characteristics to compare detector designs and methodical choices (\autoref{sec:taxonomy-final}). 
    \item \textbf{Evidence Assessment.} We assess the evidence of detection studies by examining the evaluation quality, reporting practices, and a reproducibility study of selected papers to evaluate the validity of the results and the state of published artifacts (\autoref{sec:performance-evaluation}, \autoref{sec:deployandoperate}, \autoref{sec:rep}). 
    \item \textbf{Research agenda.} We derive methodological gaps, open problems, and formulate recommendations for future web tracker detection research (\autoref{sec:sys}, \autoref{sec:discussion}). 
\end{itemize}

\section{Preliminaries}\label{sec:prelim}
We introduce relevant web tracking concepts to provide a common understanding for the rest of the paper (\autoref{sec:terminology}). 
Thereafter, we identify and clarify the primary stakeholders involved in web tracker detection (\autoref{sec:stakeholders}). 

\subsection{Terminology}\label{sec:terminology}

\begin{figure*}[t]
    \centering
    \includegraphics[width=0.89\textwidth]{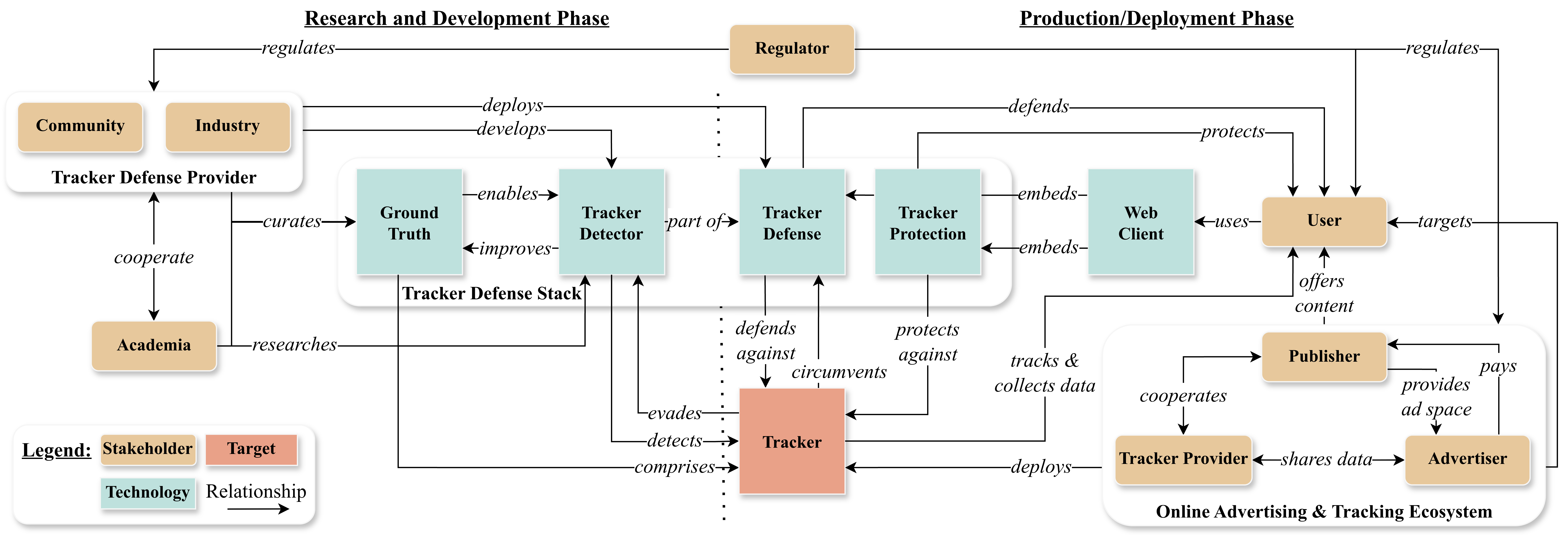}
    \caption{Various stakeholders participate throughout the life cycle of a detector. Whereas the left side focuses on detector research and development, the right side shows which stakeholders affect or are affected by detectors in the real-world.}
    \label{fig:stakeholders}
\end{figure*}

\noindent\textbf{Web tracking and tracker.} Web tracking is defined as the process of observing a user and their interactions on websites and across the web, e.g., for ads or web analytics. The entity observing is defined as the tracker that uses techniques and technologies to: identify and render users or their interactions observable; and leak information. This definition is closely related to Roesner et al.~\cite{roesner} and Cozza et al~\cite{cozza}. Other definitions use a tracker's presence in filter lists~\cite{englehardt} or denote the technique and technology itself as a tracker~\cite{gomer}. However, we abstain from adopting this narrower definition as the observational property is more generally applicable. Unless explicitly stated otherwise, we restrict this definition to browser-based web contexts and exclude non-browser environments (e.g., mobile apps).

\noindent\textbf{First- and third-party.} The \textit{first-party} is the website owner (publisher) of the visited website by the user. Websites can embed resources from an external party -- \textit{third-party}. There is a many-to-many relationship between first- and third-parties. However, the first-party and an external web server may be operated by the same legal entity. Identifying the legal entity that operates a certain web server may not be reliably possible as this information is either not public or obfuscated. Hence, researchers differentiate between first- and third-party connections considering the second-level domain. 
Therefore, the tracker as an observing entity is not a legal entity unless organizational criteria are considered. The selection of third-parties must contribute an added value to the publisher (see \autoref{sec:stakeholders} and \autoref{sec:threat-modeling}).

\noindent\textbf{Detection, defense, and protection.} Combating web tracking in general involves three approaches: \textit{detection}, \textit{defense}, and \textit{protection}. Detection concerns itself with the identification of trackers~\cite{khaleesi-2022, siby-2022, cookiegraph-2023, lee_adflush_2024}.  
Defense describes measures where the user is actively being tracked and thus uses, e.g., ad and tracker blockers with filter lists to identify and block any communication to trackers~\cite{privacybadger, uBlock, uBlockOrigin, ghostery, adblock, abp, disconnect}. Protection involves measures that shield the user from being in a tracking situation by, e.g., blocking third-party cookies or fingerprinting countermeasures~\cite{ITP, farbling}. These protection measures are usually implemented and provided by browser vendors~\cite{brave, safari, firefox, chrome}. 

The main challenge in detection is to decide when user identification, data collection, and data leakage is \textit{benign} or malicious~\cite{Krishnamurthy2006Footprint, fpdetective, Papadopoulos2019CookieSync, acar-measurement, Fouad2018MissedBF}.  
A common assumption is that techniques and technologies are harmful if users are not aware or do not benefit from it. 
User identification can be benign when authentication and authorization contribute to a secure and private environment.
Data collection can lead to more relevant ads but depending on the user it can be benign or not. 
Data leakage is inevitable as websites: may provide useful and context-dependent information to the client (location); and must adapt to its technical constraints (dimensions and orientation of the screen). 

\noindent\textbf{Identification mechanisms.}
We have \textit{stateful} and \textit{stateless} mechanisms in tracking. The former stores a local identifier as a random value on the client's side (state is altered).  
In the latter, the tracker aims to recognize the client with available or leaked information and generates an identifier at any time in a communication sequence. Hence, it is not mandatory to persist it on the client side (state remains unmodified). These mechanisms are known as fingerprints~\cite{nikiforakis_2013, laperdrix_2016}. 

\noindent\textbf{Evasion and circumvention.} Against these protection mechanisms, trackers counter with \textit{evasion} and \textit{circumvention} strategies. The conceptual difference between the two is: evasion strategies aim to pass the detection process, i.e., the detector classifies the observation process as benign, while in applied circumvention strategies, the observation is not subject to the detection process at all. 

\subsection{Stakeholders in the Detection Ecosystem}\label{sec:stakeholders}
Detection is not merely a technical challenge as it involves and affects a range of stakeholders throughout its life cycle. To better situate their roles and interdependencies, we extend the known depiction of the online advertising and tracking ecosystem~\cite{cook, chua, gomer}. Figure~\ref{fig:stakeholders} shows a high-level representation of the web tracking ecosystem through the lens of detectors.  
We abstract and simplify the ad and tracking ecosystem as the concrete inner workings are out-of-scope for this work and were already analyzed in multiple studies~\cite{cook, chua, gomer}. 

A detector undergoes two phases that separates detector and defense: research and development (training and testing); and production (deployment). The initial phase is collaborative, relying on \textit{academia}, \textit{industry} (inc. organizations), and the \textit{community} to work on detectors, defense tools, and ground truths. In addition, we identify \textit{tracker defense providers} who develop and maintain tracker defense tools. The collaboration between academia and industry ensures the refinement of detectors and addresses the dynamic, adversarial nature of web tracking -- trackers regularly adapt to evade detection and circumvent defense. 

The second phase is characterized by the \textit{users}, \textit{trackers}, the \textit{online advertising 
\& tracking ecosystem} (sharing of user data and IDs), and deployed detectors and defenses. A user's access to the web is enabled by a \textit{web client} -- a desktop or mobile browser; or WebViews.
Tracker defenses and protections may be available, e.g., as a browser extension or through built-in browser features and settings. 
Trackers (providers) can be either first- or third-parties with the goal of identifying and collecting information about the user on the same- or cross-site, and same- or cross-device. In addition, trackers can be deployed in multiple web artifacts (see~Table~\ref{tab:artifact-defs}) that affect their capabilities and reach (see~Table~\ref{tab:sok-mapping}). 
It is necessary to deploy technologies from the \textit{tracker defense stack}: Users have privacy concerns about web tracking, especially after being made aware of and sensitized to the extent of tracking ~\cite{melicher, weinshel, schaub-2016}. 

\section{Related Work} \label{sec:rel-work}
After our initial Google Scholar search, we looked through existing collections of SoK papers (in October 2024 and April 2025)~\cite{sok_list, sok_list_oakland}. This revealed no papers explicitly addressing our topic. In addition, there is no prior work that systematized the topic of detection as extensively as our SoK. 
We did not include research on the (empirical) evaluation of ad and tracker blockers as well as filter lists~\cite{mazel, ruffell, snyder, wills, traverso, merzdovnik}. These well-researched tools aim to defend the user against tracking and would violate our intervention criteria (see Table~\ref{table:picoc}). 
The listed tools provide an industry or community perspective. 
Lastly, we focus on detectors from the academic literature. This leaves a handful of review and survey papers that mainly address tracking and/or techniques and technologies. 

Besides surveying the field of web tracking, we can observe three general focus groups that were either considered individually or in combination: (i) web tracking as whole~\cite{mayer, ermakova, sok-web-tracking}, (ii) techniques and technologies~\cite{Bujlow, sanchez-rola, Sim, sok-web-tracking, fp-survey}, and (iii) tracking defense and privacy protection mechanisms~\cite{Bujlow, ermakova, Sim, abdulaziz, sok-web-tracking, fp-survey}. 
Each survey extensively addressed these focus groups in-depth over many years, with the most recent study from 2025~\cite{sok-web-tracking}. 
A comprehensive review of these focus groups in addition to detection is therefore considered out-of-scope for this SoK. The above-mentioned references and Table~\ref{tab:sok-mapping} serve as the primary starting point for these research subjects. 

Only four reviews include the topic of detection separately: as either brief overview~\cite{sanchez-rola, Sim, sok-web-tracking}; or further analysis as in Abdulaziz et al. that selected more ML-based detectors~\cite{abdulaziz}. 
However, their primary focus was on tracking defense, thus tailoring the evaluation of detection approaches to this objective. This focus is reflected in their search matrix, where detection was not used as a search term in contrast to protection, or defending -- detectors were reviewed under the umbrella of tracking defense. Furthermore, the review does not go beyond a summarization of detectors. 
To differentiate our work, we do a comparison in Table~\ref{table:rel-work-comparison}. 

\begin{table}[ht]
\caption{Related Works Comparison}
\centering
\scriptsize
    \begin{tabularx}{\columnwidth}{rcc}
        \toprule
        \textbf{Criteria} & \textbf{Abdulaziz and Frikha~\cite{abdulaziz}} & \textbf{Our Paper} \\
        \midrule
        \textbf{Focus} & Web tracking \& defense & Detectors \\
        \textbf{Search Time Period} & 2016-2021 & 1990-2025 \\
        \textbf{Type of review} & Literature review & Systematic review \\ 
        \textbf{\#Detectors identified }& $8^\dagger$ & 59 (+16) \\
        \textbf{Industry Solutions} & \CIRCLE & \LEFTcircle \\
        \textbf{Systematization} & \xmark & \cmark \\
        \textbf{Taxonomy} & \cmark\ (Tracker Defense) & \cmark\ (Detectors) \\
        \textbf{Qualitative analysis }& \xmark & \cmark \\
        \textbf{Reproducing studies} & \xmark & \cmark \\
        \textbf{Artifacts} & \xmark & \cmark \\ 
        \bottomrule
    \end{tabularx}
    \begin{minipage}{\columnwidth}
    \vspace{1mm}
    \centering
    \scriptsize
    \textbf{Legend:} \cmark\ criterion is met; \xmark\ criterion is not met; \CIRCLE\ inclusion; \LEFTcircle\ partial inclusion. \\
    $\dagger$ out of $n=30$ papers reviewed, only eight explicitly developed a detector. 
    \end{minipage}
\label{table:rel-work-comparison}
\end{table}

Lastly, the methodology of the review papers is often unclear. Methodological approaches to summarize and synthesize the knowledge in web tracker research were rarely reported, except for~\cite{ermakova, abdulaziz} in varying degrees. This does not degrade the quality and results as we can only assess reporting and not methodological quality. However, it leaves questions regarding the reproducibility, completeness, and potential biases. Therefore, we provide a usable methodology for other researchers in the following section. 

\section{Research Methodology}\label{sec:method}

We introduce our five-stage methodology to systematize web tracker detection research: (i) a \textit{systematic review} to identify relevant literature on our topic (\autoref{sec:sys-review}), (ii) \textit{systematization} as a qualitative approach, (iii) a \textit{taxonomy building framework} from information systems research (\autoref{sec:taxonomy}), 
(iv) \textit{reproducibility experiments} to verify whether we can reproduce the results reported in selected studies (\autoref{sec:repro}), and (v) an assessment of the quality of evidence (\autoref{sec:quality}). 

\subsection{Systematic Review}\label{sec:sys-review}
\begin{figure}
    \centering
    \includegraphics[width=0.8\columnwidth]{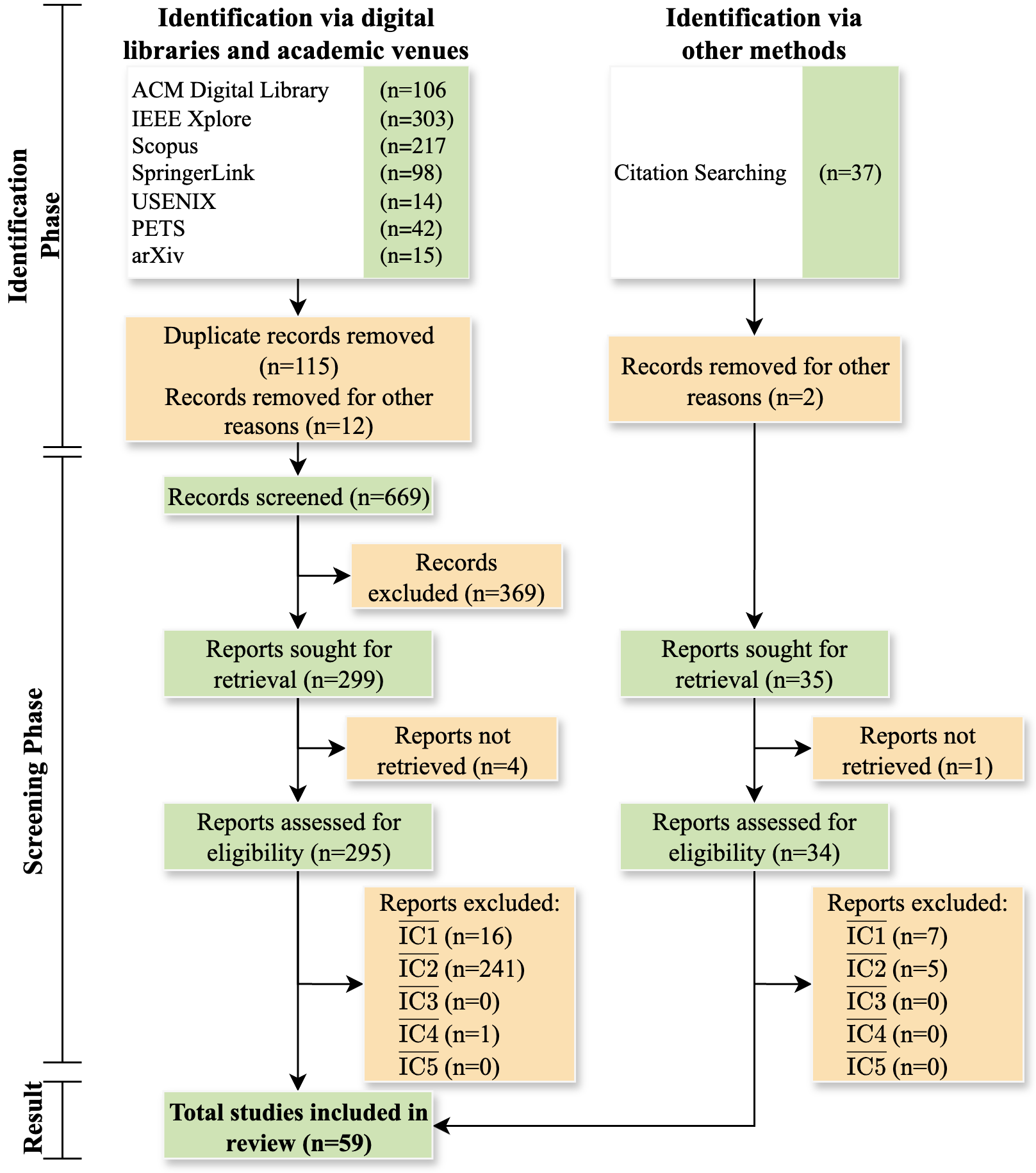}
    \caption{Adapted PRISMA flow diagram~\cite{prisma}. 
    }
    \label{fig:review-overview}
\end{figure}

We follow the Preferred Reporting Items for Systematic reviews and Meta-Analyses (PRISMA) framework, which originated from medical and healthcare research~\cite{prisma}. PRISMA provides a standardized guideline for documenting literature identification, screening, eligibility assessment, study inclusion, and reporting in systematic reviews and meta-analyses. We adopt the PRISMA items most relevant to our SoK, while omitting items that are not well suited with the broader synthesis and systematization goals of a SoK. This ensures a transparent, reproducible search, and validity of our findings. Identifying almost all available records allows us to formulate generalizable conclusions that reflect the current literature~\cite{mulrow}. Our search process is split in two rounds (Figure~\ref{fig:review-overview}). The first round includes literature from digital libraries. Relevant studies might be missing from the first round as digital libraries are not indexing the whole corpus of studies. Therefore, a second round applies citation searching to the identified studies from the first round. 

\noindent\textbf{Data Sources.}
We select our studies from two information sources: digital libraries (ACM Digital Library~\cite{ACMDigitalLibrary}, IEEE Xplore~\cite{IEEEXplore}, Scopus~\cite{Scopus}, SpringerLink~\cite{SpringerLink}, arXiv~\cite{arXiv}) and relevant security and privacy venues that are not fully covered by the digital libraries (USENIX~\cite{USENIXProceedings}, PETS~\cite{PoPETs}). 
Furthermore, we use Google Scholar for forward and backward citation searching combined with ACM Digital Library and IEEE Xplore as recommended by~\cite{gusenbauer}. 

\noindent\textbf{Search Strategy.}
We first curate a list of search terms in five steps: (i) listing and defining keywords from our experience, (ii) extracting all keywords from past research studies on web tracking, (iii) reducing the number of keywords through abstraction and relevance, (iv) developing a base search matrix, and (v) by iteratively adjusting the search matrix for each data source to identify a sufficient number of studies (search matrices are given in Appendix~\ref{app:sys-review}). 

\definecolor{snet-red}{rgb}{0.772, 0.059, 0.122}

In addition, filters were applied whenever applicable: time frame (1989–present), language (English), and the type of publication (conference or journal). We conducted our search and exported the results between October 25\textsuperscript{th} and November 8\textsuperscript{th}, 2024. From the corpus of records, we remove duplicates through automated means by defining \textit{Regular Expressions} that compare the title and DOI. Then we remove records that are not reports, e.g., introductions, forewords, mini-track descriptions, or workshop reports.

\noindent\textbf{Screening Strategy.}
To minimize bias and to ensure consistency in our screening process, we assign two experts to independently screen a paper. The compiled review reports are then compared and discussed. Each decision is accompanied by a third reviewer and disagreements were solved through consensus or by said reviewer~\cite{prisma, kitchenham}. 

In the first screening phase, we screen a record's title, abstract, and keywords with the goal of filtering out topics such as e-learning, e-voting, or fingerprinting for authentication and law enforcement. 
In the second screening phase, we assess the whole paper according to the pre-defined eligibility criteria in Table~\ref{table:eligibility-criteria}. The assessment of a paper satisfying the inclusion criterion (IC) is terminated when the logical complement of one $\overline{\text{IC}}$ is met. To assess the level of agreement between both reviewers, we calculate the inter-rater reliability using Krippendorff's alpha~\cite{krippendorf}, resulting in a score of $\alpha=0.801$ (indicating a reliable level of agreement). Although our percent agreement is higher, Krippendorff's alpha results in a lower value due to the high number of negative samples, i.e., excluded reports~\cite{geijer}. Thus, we calculate Gwet's AC\textsubscript{1} to account for the imbalance, resulting in an inter-reliability score of $0.943$~\cite{gwet}. 

\begin{table}[ht]
    \caption{Criteria for the eligibility assessment of records.}
    \centering
    \scriptsize
	\begin{tabularx}{\columnwidth}{rX}
            \toprule
		\textbf{Identifier} & \textbf{Eligibility Criteria} \\
		\midrule
            \textbf{IC1} & \textbf{Study Field:} The study must have conducted research in the field of web privacy in the context of web tracking. \\ 
            \textbf{IC2} & \textbf{Study Design:} The study must have developed a detector (this excludes solely identification approaches for web tracker/web tracking as well) and evaluated using empirical, such as quantitative, qualitative, or mixed methods. \\ 
            \textbf{IC3} & \textbf{Detector Design:} The detector must be either based on heuristics or a model from machine learning (ML), deep learning (DL), natural language processing (NLP), or a combination thereof.  \\ 
            \textbf{IC4} & \textbf{Dataset Specification:} The study has to explain how and when the data was collected or if an existing dataset has been used, a reference included. The employed ground truth has to be comprehensible. \\ 
            \textbf{IC5} & \textbf{Study Language:} The study must be written in English. \\ 
		\bottomrule
	\end{tabularx}
 \label{table:eligibility-criteria}
\end{table}

\noindent\textbf{Citation Searching.}
Forward and backward citation searching using the same eligibility criteria was conducted between January 2\textsuperscript{nd} and January 10\textsuperscript{th}, 2025. This has been shown to be a valuable supplementary method~\cite{hirt}. 

\subsection{Web Tracker Detection Taxonomy}\label{sec:taxonomy}
We develop the taxonomy (\autoref{sec:taxonomy-final}) through multiple iterations following the methods by Nickerson et al.~\cite{nickerson} and Kundisch et al.~\cite{kundisch}. This ensures methodological rigor and transparency. 
Based on Nickerson et al.'s recommendation and our scope (\autoref{sec:intro}, Table~\ref{table:picoc}), we define our meta-characteristic: \textit{The technical-, methodological-, and operational-attributes of detectors}. We will start with a conceptual-to-empirical approach (deductive) based on the existing theory and our own expertise. Then we refine the taxonomy with an empirical-to-conceptual approach (inductive) based on concrete detectors from our systematic review. Researchers may benefit by: having a new conceptualization of detectors, situating their future work, and identifying potentially new approaches. Community and industry may benefit in the selection of detectors for their products, e.g., to strengthen the privacy defense of browsers. 

\subsection{Reproducibility Study}\label{sec:repro}
Numerous academic fields in the last decade have suffered from reproducibility problems of scientific outcomes~\cite{nature-reproducibility}. Repeatability, reproducibility, and replicability~\cite{ArtifactReviewBadging} are important for the independent verification of a study's methods and results. 
This importance is exemplified by Tier 1 security and privacy conferences requiring artifacts. 
Following Table~\ref{table:eligibility-criteria}, all reviewed studies produced at least one artifact: either a dataset and/or a detector. 

A reproducibility study has not been done in this field. However, reproducibility has been conducted as either artifact reviews~\cite{siby-2022, shuang_dumviri_2025} or verifying the results using past detectors as a baseline ~\cite{khaleesi-2022, rieder_beyond_2025, cookiegraph-2023}. Following our objectives, RQ1, and RQ3, we systematize the methodologies and artifacts among other things in conjunction with performing reproducibility experiments. Doing so strengthens our answer to RQ4 by highlighting limitations and gaps for future research. However, performing a full reproducibility study such as~\cite{tier1-reproducibility} would be out-of-scope for a SoK paper. 

Therefore, we conduct limited reproducibility experiments of selected studies (\autoref{sec:rep}). Combined with the overall systematization, we clarify the current state of detector reproducibility and existing challenges. 

\subsection{Assessing the Quality of Evidence}\label{sec:quality}
Systematic reviews depend on the reliability of the included studies -- omitting quality assessments risks increasing biases or errors. Some studies might use non-representative or skewed datasets; or inadequate validation; or unavailable artifacts. These may lead to unreliable results. Assessing the quality of studies improves the precision of synthesized findings and mitigates misleading conclusions. This enhances the review’s credibility~\cite{kitchenham, grade, cochran}. 

Therefore, we assess the quality of evidence of studies by means of a framework (Appendix~\ref{sec:quality-results}) that only focuses on the evaluation quality (of the detector) and not the overall quality of the study. Following PRISMA and the Cochran Handbook, we select the Grades of Recommendation, Assessment, Development and Evaluation (GRADE) approach~\cite{grade} and adapt it to the scope of this SoK paper. 

\section{Results of the Search and Selection Process}\label{sec:search-results}
A total of 59 primary studies were identified in our systematic review (conference (n=43); journal (n=16)) as shown in Figure~\ref{fig:review-overview}. 
There are four groups of studies that were excluded in the second screening phase: (i) not having developed a detector; (ii) not having evaluated a detector; (iii) using detectors for different purposes; and (iv) ad detectors. 
An example for group (iii) would be assessing the tracker prevalence in web privacy measurements~\cite{libert, mayer, englehardt, acar-measurement, papadogiannakis-2021}.
The reasoning behind group (iv) is that the primary target of ad detectors are ads instead of WebTs. However, an embedded third-party script can have a dual-purpose and ad-intentioned web elements can support the tracking of users~\cite{siby-2022, ikram, bashir}. 
Groups (iii) and (iv) are still relevant for a holistic picture of the field, but cannot be adequately and fairly compared to primary studies. Therefore, we will label them as supplementary studies (marked as, e.g., \cite{orr}\textsuperscript{$\ddag$}). 

Applying our eligibility criteria shows that all detectors in the final review use either heuristics, ML, deep learning, or natural language processing (IC3). Moreover, all studies described dataset specifications (methodological reporting standard), except for \cite{fpflow-2022}\textsuperscript{$\ddag$}. The low number of reports that met $\overline{\text{IC1}}$ indicates that our first screening step was successful in removing studies unrelated to web tracking. Nevertheless, our first searches still missed 22 primary studies that were then identified through citation searching. 

\noindent\textbf{Limitations of the Screening.}
The first limitation is the strict inclusion criterion. IC2 might be misrepresenting due to the aforementioned grouping issue and could be split into multiple IC. 
The second limitation is the order of the inclusion criterion. For example, ordering IC3 before IC2 would have shown additional studies, even if they were excluded later due to missing empirical evaluations. Addressing both limitations would reflect the literature more accurately. 

\noindent\textbf{Bias Assessment.}
The last limitation surrounds the topic of biases that are always part of reviews -- they can only be reduced but not eliminated~\cite{cook2008}. We address the following biases that might have influenced our results:
(i) a \textit{Language Bias} as non-English publications were excluded that may contain detectors; 
(ii) a \textit{Publication Bias} as we mainly select published peer-reviewed studies and limited the selection of studies on arXiv, along with the selection of only high-ranked venues as additional data sources; 
(iii) an \textit{Evidence Selection Bias} as we excluded old studies on arXiv;  
(iv) a \textit{Search Bias} that should have been prevented with the selection of eight different data sources and the inclusion of the majority of eligible studies~\cite{ewald};  
(v) a \textit{Selection Bias} due to exclusion of most gray literature~\cite{paez} and IC3, resulting in the omission of filter lists, ad- and tracker blockers, and measurement studies. 

\section{Taxonomy}\label{sec:taxonomy-final}
Before we dive into the systematization, we introduce our taxonomy on detectors as the overall categorization of the literature. 
Past review and survey papers focused and proposed taxonomies on techniques and technologies (\autoref{sec:rel-work}). Papers about detectors introduced broadly-defined categories that are based on: (i) the learning approach~\cite{annamalai-2024, lee_adflush_2024}; (ii) the feature-level~\cite{khaleesi-2022, rieder_beyond_2025, bytecode-2023}; (iii) the resource-level~\cite{kargaran-2021, lee_net-track_2023}; or (iv) in the context of techniques and technologies~\cite{raschke_tex-graph_2023}. These categories represent individual parts of a full taxonomy that has never been proposed before. 

\begin{figure}[t]
    \centering
    \includegraphics[width=0.80\columnwidth]{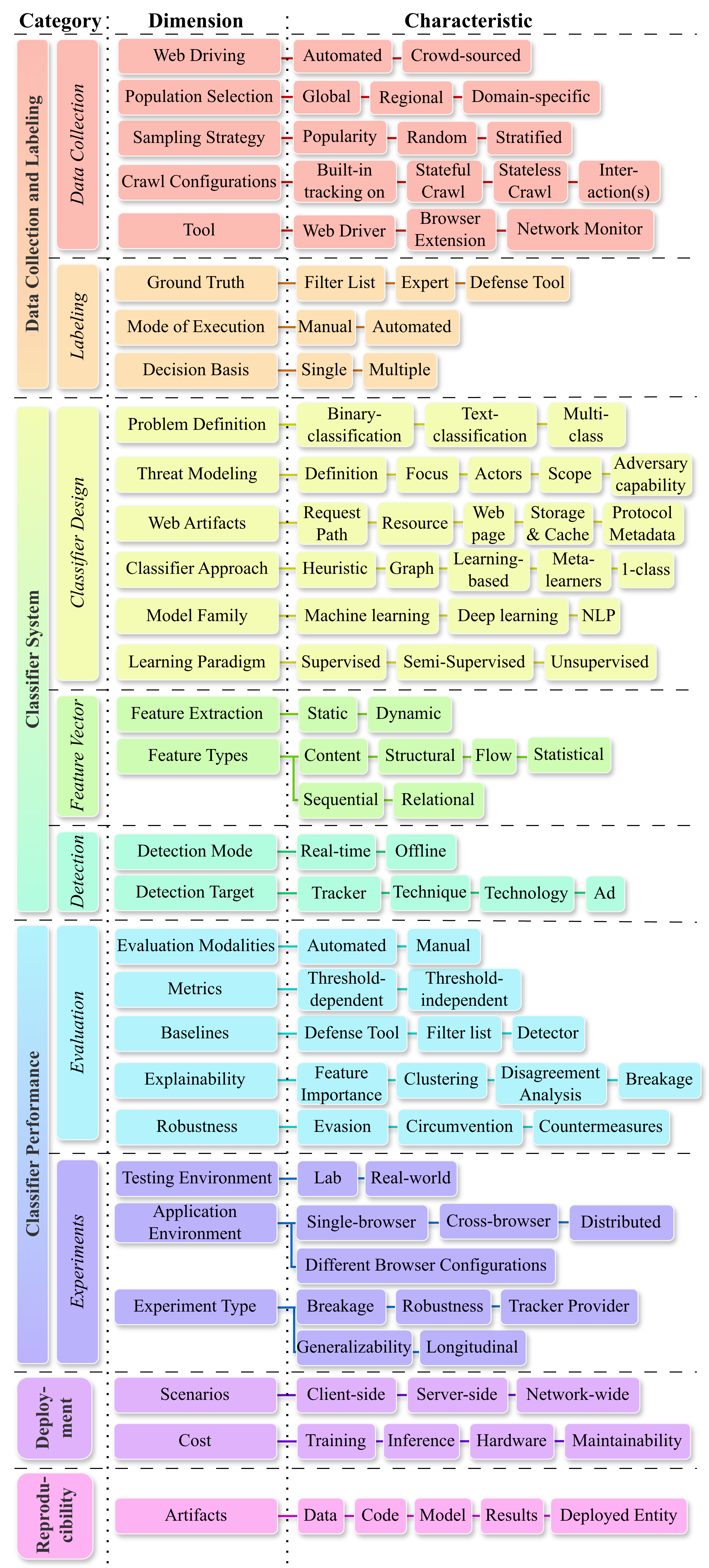} 
    \caption{The detector taxonomy comprising ten categories with their related dimensions and characteristics. A detector can fulfill more than one characteristic per dimension. The color gradient represents the flow of information of data from higher entropy to lower entropy~\cite{shannon}.} 
    \label{fig:taxonomy}
\end{figure}

In the context of the \textit{tracker defense stack} (Figure~\ref{fig:stakeholders}), the detector is part of tracker defense where it classifies a given input according to a pre-defined detection target. 
For example, the input may be a network request or a web resource, while the detection target may be whether the input is tracking-related or benign. 
A detector takes as input a labeled dataset $\mathcal{D}=\{(a_i,y_i)\}_{i=1}^n$ where $n\in\mathbb{N}$, $a_i\in\mathcal{A}$ the crawled records, and $y_i\in\mathcal{Y}$ the set of labels based on a ground truth. 
Then $\mathcal{D}$ is transformed via a feature mapping function $\phi: \mathcal{A}\to\mathcal{X}$, where $\mathcal{X} \subseteq \mathbb{R}^d$ is the set of features. 
For instance, if $a_i$ is a network request, $\phi(a_i)$ may encode features such as URL structure or request parameters. 
The detector is then defined as a composed mapping of a parameterized feature encoder $\lambda$ (function of $x_i \in \mathcal{X}$) with a probabilistic scoring function $f$ where $f:\mathbb{R}^{d}\to[0,1]$, $\lambda \circ f$. The output is then a score $s$ (estimated likelihood that the input is tracking-related) and a decision $\delta_{\tau}(s)=[s\ge\tau]\in\{0,1\}$ for threshold $\tau \in [0,1]$. 

Based on our methodological approach (\autoref{sec:taxonomy}) and our definition of a detector, we use a typical ML-workflow as the structural basis for our taxonomy (Figure~\ref{fig:taxonomy}) and our systematization (\autoref{sec:sys}).
We observed most detectors can be either mapped onto its components or explicitly follow this workflow. 
The taxonomy abstracts over implementation details while preserving the essential parts that determine detector design and performance. This enables a consistent comparison across heterogeneous approaches. 

\section{Systematization}\label{sec:sys}
Building on the taxonomy, we use it to structure for the following systematization. Our primary approach is to apply a meta-science perspective, i.e., identifying patterns across detection studies and how they produce evidence, to then systematize the research on detection. This approach reflects evidence-based practices from medical research. 
Due to the large number of references, we will mostly give examples and provide supplementary information in Table~\ref{tab:tracker-detector-table}. 
Inspired by Wei et al.~\cite{solk}, we label our following statements according to trends \trend{\textbf{T\#}}, limitations \limitation{\textbf{L\#}}, recommendations \recommendation{\textbf{R\#}}, and future work opportunities \opportunity{\textbf{O\#}}. We provide an overview of all labeled statements in Appendix~\ref{app:labels} (Table~\ref{table:summary-labels}). 

\definecolor{snet-green}{HTML}{0cc5b3}

\subsection{Threat Modeling}\label{sec:threat-modeling}
Threat modeling is a common method in (web) security research as it helps to outline the actors, threats, and mitigations of the system~\cite{Drake2025ThreatModeling}. The two main goals of detection are: detecting trackers, tracking, and advertising (providers are the adversaries); and enhancing user privacy. Adversaries are  involved in anti-ad- and tracker blocking~\cite{anti-adblock}, leaving detectors under threat, either in the environment they are located in or the detector itself. 

\noindent\textbf{First seen in 2022, threat modeling is a recent addition to detection research~\cite{siby-2022, cookiegraph-2023, munir_purl_2024, amjad-2024, polcak_jshelter_2023} \trend{\textbf{T1}}.} 
Threat models are not only recent, but also heterogeneous in modeled objects, actors, trust assumptions, and adversary capabilities. Among the set of detection studies with explicit threat models, they support their claims but differ in what they model.  
\textsc{WebGraph}\cite{siby-2022} centers on adversarial evasion of its detector and models an adaptive third-party tracker with limited first-party cooperation and concrete inference-time manipulations. \textsc{CookieGraph}\cite{cookiegraph-2023} and \textsc{PURL}\cite{munir_purl_2024} focus on tracking techniques (first-party cookie tracking and link-decoration-based identifier sharing), centering on a well-scoped ecosystem and its actors with explicit assumptions about users, publishers (not \textsc{PURL}), and adversaries. \textsc{NoT.Js}\cite{amjad-2024} differs by adopting a breakage-centric perspective and modeling mixed scripts and the conditions under which coarse-grained blocking either misses tracking or breaks legitimate functionality. \textsc{JShelter}\cite{polcak_jshelter_2023} is broader and frames threats at the browser-protection level for users. Consequently, these threat models support qualitatively different conclusions, ranging from robustness; to tracking-techniques; to breakage; and to browser protection. 

Lastly, the majority of studies since 2022 (n=24) do not include a threat model, thus leaving open questions about potential vulnerabilities or evasion and circumvention \opportunity{\textbf{O1}}. Some studies outline and discuss parts of threat modeling, e.g., robustness and how an adversary could evade or circumvent a proposed detector~\cite{yu, khaleesi-2022, lee_adflush_2024, bytecode-2023, rieder_beyond_2025}. 

\noindent\textbf{Researchers should define threat models to clarify a detector's scope and assumptions, and to support evaluations and conclusions \recommendation{\textbf{R1}}.} The deployed detectors operate in an adversarially adaptive environment~\cite{anti-adblock, snyder, dao-cname-journal}. We do not know empirically the extent of an adversary's evasion or circumvention, or model manipulations in the real-world (as detectors are not widely deployed and used by users). 
Threat models in detection provide context and delineate which conclusions about robustness, breakage, deployability, and generalizability are justified. Researchers should consider possible threats and vulnerabilities for the detection and the detector in their studies \opportunity{\textbf{O2}}. To support this, we provide several analyses, such as \autoref{sec:stakeholders} to identify actors and trust levels or a mapping between a detector's used web artifact(s) and \autoref{sec:classifier-design} to assess an adversary's attack surface and capabilities. Furthermore, researchers can use  frameworks such as by OWASP~\cite{Drake2025ThreatModeling}.

\subsection{Data Collection}\label{sec:data-collection}
Datasets play an integral part in detection research to train and test proposed classifiers as well as to identify and measure web tracking. Every study collects a new or reuses a dataset. The data collection follows a similar methodological approach across studies, where only two reported insufficiently in comparison~\cite{adremover, cheng_using_2022}. However, when examining criteria for web measurement studies by Demir et al.~\cite{repro-web-measurement-studies}, the reporting and crawling are insufficient and should be more carefully planned in the future \recommendation{\textbf{R2}}. 

\noindent\textbf{Data collection increased in scale and leveraged the popularity-based top sites ranking lists \textsc{Alexa}~\cite{alexa} and \textsc{Tranco}~\cite{tranco} to identify a suitable sample \trend{\textbf{T2}}.} 
The number of crawled web pages increased over the years, from a few hundred in the beginning~\cite{orr, bhagavatula}\textsuperscript{$\ddag$} to tens of thousands~\cite{lee_adflush_2024, shuang_dumviri_2025} or even millions today~\cite{campeny-2024, ast-trans}. For that, researchers use crawlers that visit a number of websites from multiple categories according to popularity-based ranking lists (\textit{sampling strategy}) or visit websites from one specific domain~\cite{kaizer},~\cite{moti}\textsuperscript{$\ddag$}. Earlier works often leveraged \textsc{Alexa}'s top sites, while newer research has shifted to \textsc{Tranco} as \textsc{Alexa} was discontinued and found to be biased~\cite{tranco, ranking-lists-comparison}. There are a few exceptions of other used lists such as \textsc{Quantcast}~\cite{quantcast, bau}, \textsc{Majestic}~\cite{majestic, meyer_detecting_2024}, and \textsc{CrUX}~\cite{crux, annamalai-2024}. Although this popularity-based sampling strategy is adopted by most, a few studies combine strategies: popularity and random~\cite{dao}; popularity and stratification~\cite{annamalai-2024}; or all three combined~\cite{iqbal-fingerprinting-2021, siby-2022, munir_purl_2024}. These strategies can also be applied to the resulting dataset from the crawl~\cite{campeny-2024}. 

\noindent\textbf{Popularity-based ranking lists may not result in a representative dataset on web tracking \opportunity{O3}.}
An assumption that underlies ranking lists is the representativeness of the dataset -- whether the collected data reflects the true distribution of web tracking (\textit{sampling bias}). The scope of most studies is constrained by an overreliance on the most popular websites (10K)~\cite{kargaran-2021, khaleesi-2022}, and the first 10K websites might not be representative of the web's diversity. However, the top 10K reflect the true tracker distribution (imbalance of trackers and non-trackers) more accurately~\cite{rieder_beyond_2025, adgraph} than the top 5K (almost balanced distribution)~\cite{shuang_dumviri_2025}. This entails a more reliable base rate for training. 
Still, biases arise: popularity is defined differently; and countries, client platforms, and website categories are not represented equally~\cite{ranking-lists-comparison}. 
Future studies should understand, reflect, and mitigate these biases as they are influencing detectors \recommendation{\textbf{R3}}. Furthermore, it neglects the long-tail of less-visited websites where niche trackers or region-specific tracking might occur. This could be mitigated by a stratified random sampling approach of different ranges such as 1K-10K and 10K-100K \recommendation{\textbf{R4}}. 
Another form of representativeness and diversity is related to the human-likeliness of automated crawls as browsing involving real users can reveal more fingerprinting~\cite{annamalai-2024, fp-real-world}. 
Crowd-sourced datasets~\cite{rizzo-unveiling-2021, yu} could be used but crawling these faces several challenges such as privacy and ethical requirements, recruitment challenges, and costs. Using Generative Adversarial Networks to create a synthetic dataset may be of help but first requires such a crowd-sourced dataset for training and potential adaptions over time as the web constantly changes. 

\noindent\textbf{Prior work predominantly uses \textsc{Selenium} and \textsc{OpenWPM} with different configurations to automate the crawling of today's desktop websites with \textsc{Firefox} \trend{\textbf{T3}}.}
The large-scale web privacy measurements can be traced back to two developments: more powerful computational hardware (inc. cloud); and software such as \textsc{Selenium}~\cite{selenium} and in particular \textsc{OpenWPM} in 2016~\cite{englehardt}. These helped in standardizing how data is collected through controlled and automated crawling. In general, researchers either developed either: their own client-based or proxy-based crawler (usually based on \textsc{Selenium}); or used existing ones (with custom extensions or proxies) such as \textsc{OpenWPM}, \textsc{FourthParty}~\cite{mayer}, \textsc{T.EX}~\cite{t.ex}, \textsc{WTPatrol}~\cite{yang}, \textsc{Puppeteer}~\cite{puppeteer}, or \textsc{OmniCrawl}~\cite{cassel} as part of a browser-extension, a web driver (usually in headless mode), or an instrumented browser. This led to a range of different crawlers and crawler versions, thus affecting reproducibility and accurate measurements~\cite{webrec} \limitation{\textbf{L1}}. We refer to~\cite{webrec, krawlers, crawlers-impact} for an in-depth analysis of crawlers and crawling. From the literature, we can further observe a \textit{browser bias}, i.e., most studies rely on the instrumentation of a \textsc{Firefox} instance that was primarily driven by \textsc{OpenWPM}. Although other crawlers support additional browser drivers~\cite{t.ex, cassel}, the unspoken good practice of using \textsc{Firefox} (that has a relatively small user-base~\cite{statcounter}) solidifies its continued usage. 
One reason for its preference lies in its adaptability and extensibility compared to \textsc{Chromium}, facilitating static (fetching what is loaded at runtime) and dynamic (capturing resources and interactions during runtime) measurements when visiting a web site. 

\noindent\textbf{Detection research lacks a community-standard benchmark suite that combines shared datasets, metrics, and a reference crawling testbed to facilitate comparability, reproducibility, and external validity \opportunity{\textbf{O4}}.}
Every study collects a new dataset on client-side observable web tracking instead of solely reusing or reproducing existing ones.
Studies employ various crawling configurations such as: only visiting landing pages, simulating basic user interaction via clicks, one geographical location, disabled tracking protection, or timeouts. This demonstrates the need for a proven and standardized crawling testbed or configuration set.

\subsection{Data Labeling}\label{sec:data-labeling}
Detection has predominantly been formulated as a classification problem, requiring a labeled dataset and a ground-truth (usually deterministically via a filter list or heuristics) in pre- or post-training. Thus the performance and real-world applicability of detectors is directly affected. 

\begin{table}[ht]
    \centering
    \caption{Alignment of used Filter Lists and Web Tracking Definition.}
    \ssmall
	\begin{tabularx}{\columnwidth}{rcccc}
            \toprule
		\textbf{Filter List} & \textbf{Target} & \textbf{Match} & \textbf{Kept} & \textbf{FP} 
            \\
		\midrule
            \textbf{\textsc{EasyList}}~\cite{easyList} 
            & Ads 
            & \CIRCLE 
            & \cmark 
            & \Circle 
            \\ 
            \textbf{\textsc{EasyPrivacy}}~\cite{easyPrivacy} 
            & Tracking
            & \CIRCLE
            & \cmark 
            & \CIRCLE 
            \\    
            \textbf{\textsc{Disconnect.me}}~\cite{disconnect} 
            & Tracking
            & \CIRCLE 
            & \cmark 
            & \CIRCLE 
            \\ 
            \textbf{\textsc{Peter Lowe}}~\cite{peterlowe} 
            & Ad servers
            & \CIRCLE 
            & \cmark 
            & \Circle 
            \\ 
            \textbf{\textsc{Blockzilla}}~\cite{blockzilla} 
            & Ads \& Tracking
            & \CIRCLE 
            & \xmark 
            & \Circle 
            \\ 
            \textbf{\textsc{uBlock Privacy}}~\cite{ublockprivacy} 
            & Tracking 
            & \CIRCLE 
            & \cmark 
            & \LEFTcircle 
            \\ 
            \textbf{\textsc{uBlock Filters}}~\cite{ublockfilters} 
            & Ads 
            & \CIRCLE 
            & \cmark 
            & \Circle 
            \\ 
            \textbf{\textsc{hpHosts}}~\cite{hphosts} 
            & Ads \& Tracking
            & \CIRCLE 
            & \xmark 
            & \Circle 
            \\ 
            \textbf{\textsc{WhoTracks.Me}}$^\ddagger$~\cite{whotracksme} 
            & Ads \& Tracking
            & \CIRCLE 
            & \cmark 
            & \CIRCLE 
            \\ 
            \textbf{\textsc{Fanboy's Social}}~\cite{fanboysocial} 
            & Social Media Widgets 
            & \CIRCLE 
            & \cmark 
            & \Circle 
            \\ 
            \textbf{\textsc{Fanboy's Annoyance}}~\cite{fanboyannoyance} 
            & Website Annoyances
            & \LEFTcircle 
            & \cmark 
            & \Circle 
            \\ 
            \textbf{\textsc{Warning Removal List}}~\cite{warninglist}  
            & Anti-adblock warnings
            & \Circle 
            & \cmark 
            & \Circle 
            \\ 
            \textbf{\textsc{IBM X-Force Exchange}}~\cite{ibm-platform} 
            &  Threat Intelligence
            & \Circle 
            & \cmark 
            & \Circle 
            \\ 
            \textbf{\textsc{Anti-Adblock Killer}}~\cite{anti-adblock-killer} 
            & Anti-ad blockers
            & \Circle 
            & \cmark 
            & \Circle 
            \\ 
            \textbf{\textsc{Squid Blacklist}}$^\dagger$ 
            &  General-purpose blocking
            & \Circle 
            & \xmark 
            & \Circle 
            \\
		\bottomrule
	\end{tabularx}
    \begin{minipage}{\columnwidth}
    \vspace{1mm}
    \centering
    \scriptsize
    \textbf{Legend:} Alignment strength can be: \CIRCLE\ direct or strong, \LEFTcircle\ weak, \Circle\ none or unknown. Their status can be: \cmark\ actively maintained, \xmark\ outdated or discontinued. Lists address fingerprinting (FP): \CIRCLE\ yes, \LEFTcircle\ implicitly, \Circle\ none or unknown.
    $^\dagger$We found no plausible reference for this filter list (used by~\cite{adgraph}). 
    $^\ddagger$Is not a filter list per se but rather a database. 
    \end{minipage}
 \label{table:filter-lists}
\end{table}

\noindent\textbf{The de-facto standard for ground-truths has been filter lists albeit their known limitations and moderate performance~\cite{alrizah, hashmi, snyder}} \trend{\textbf{T4}}. 
Their dominant use stems from their extensive list of labels and the trust in the experience of the community maintaining these lists. Nevertheless, they suffer from human errors, biases, noise~\cite{snyder}, and they favor functionality over truth for mixed trackers to avoid website breakage for users. Filter lists are primarily designed for practical blocking decisions rather than for producing research labels -- all of which can contribute to the model learning patterns where legitimate behavior is labeled as tracking \limitation{\textbf{L2}}. In addition, they have further limitations: simple rules are preferred as complex rules would be parsed slower; Manifest v3~\cite{BraveShieldsManifestV3} limits the flexibility of rule definitions and the size of lists; and many rules are unused. 
The most frequently used filter lists for ground truth are \textsc{EasyList} and \textsc{EasyPrivacy} with the former only partially addressing Web tracking. Although both lists are essentially URL (inc. hostname) filters, they are used to label, e.g., JavaScript code executed for fingerprinting~\cite{campeny-2024}. The granularity of rules differs, with most instances blocking communication to a certain host entirely and a minority of rules blocking specific resources~\cite{snyder}. The motivation behind the design of a rule is not always transparent, thus, detectors may adopt inconsistent labeling, which exacerbates the explainability of a classifier's decisions. To address these issues, studies manually inspect a sample of misclassifications~\cite{iqbal-fingerprinting-2021, munir_purl_2024, shuang_dumviri_2025, cccc-2021} and iteratively correct the ground truth in instances where the detector was right. Currently, future studies do not benefit from these corrections \opportunity{\textbf{O5}}. Developed heuristics and models should be used to complement common sources for ground truth \recommendation{\textbf{R5}}. 

\noindent\textbf{The labeling definition should reflect the tracker (tracking) definition by the authors \recommendation{\textbf{R6}}.}
Doing so allows the models to learn correct patterns associated with web tracking and to ensure generalizability as well as validity by classifying the correct targets. A problem arises when the tracker definition is informal and tied to the presence of a tracker in a filter list, i.e., the ground truth is not only a collection of accepted and trusted labels but also the definition of the subject to study~\cite{rieder_beyond_2025, lee_adflush_2024}. This results in a coarse-grained circular definition of tracking with limited explanatory power and a biased evaluation towards knowledge already encoded in the lists -- a model is rewarded for re-discovering what is already known \limitation{\textbf{L3}}. 
The fundamental question of what constitutes a tracker shifts to whether a hostname is on a filter list. 
In addition, every hostname not on the filter list would automatically be a non-tracker, including yet-to-be-seen trackers, which is not the case with a formal definition. 
Filter lists should be constrained to their labeling characteristic -- if not, then the connection should be made to the authors definition of a tracker in their threat modeling. To help authors in their selection, we map prior used filter lists to our definition (\autoref{sec:prelim}) in Table~\ref{table:filter-lists}. Most are tied to tracking or ads, thus matching the observational and/or data leakage property of trackers. Caution is advised with some filter lists that do not have a direct connection to tracking or that have been discontinued. 

\noindent\textbf{Labeling fingerprinting uses a mix of different methods involving manual input, as filter lists alone are insufficient \trend{\textbf{T5}}.} 
In contrast to labeling URLs or domains, fingerprinting requires more precise labeling approaches as the lines between functional and tracking resources are more blurry. Fingerprinting is frequently embedded in these types of resources, so naïve blocking can cause site-breakage, thus impacting user experience and retention. Therefore, researchers use semi-automated, iterative methodologies to generate and refine ground truths~\cite{iqbal-fingerprinting-2021, bahrami_fp-radar_2022},\cite{tracer}\textsuperscript{$\ddag$}. These can include: expert manual review/labeling of classification results/scripts; iterative re-training; behavioral-evidence; and the use of self-developed re-usable heuristics, e.g., heuristics by \cite{iqbal-fingerprinting-2021} for \cite{annamalai-2024, bahrami_fp-radar_2022}. 
A high-quality ground truth or unified labeling definition does not exist yet \opportunity{\textbf{O6}}. 

\noindent\textbf{The current paradigm of binary labels restricts detection at the cost of more precise detectors, explainability, and transparency \opportunity{\textbf{O7}}.} 
Detection primarily classifies observations into tracker and non-tracker, limiting the ability to precisely differentiate between mixed-purpose resources. This classification mode is further substantiated by filter lists such as \textsc{EasyList} and \textsc{EasyPrivacy}, which only provide information on whether a specific resource is ad- or tracking-related and whose mode of detection is also coarse-grained (domain or URL-level). When authors use filter lists to label their dataset(s), they adopt the categories of these lists, e.g., tracker or ad. These categories are broadly defined and do not differentiate within categories. Therefore, future studies could introduce more fine-granular labels or multi-labels, e.g., domain A has the labels: (i) tracking; (ii) the technique bounce tracking; and (iii) the provider Google. This could enable more precise blocking approaches~\cite{trackersift}.

\subsection{Classifier System}\label{sec:classifier-design}
The next step involves the development of a classifier system, encompassing (i) data pre-processing, (ii) feature engineering, (iii) classifier design and training, and (iv) detection (inference and action).  
Detection determines: where/when the classifier is invoked; what is labeled; and how decisions propagate under constraints of feature availability (page-local, stateless, or stateful). 
From our contextualization and analysis of the whole literature, we abstract a fundamental unit called web artifact (see Table~\ref{tab:artifact-defs}). Features are extracted from this unit and a classifier assigns a label (classifier target) to it. 
It is not equivalent to detection targets that refer to tangible objects such as trackers, advertisements, the activity of web tracking, allow-listed entities, or mixed trackers. Based on web artifacts, we can further map them to techniques and technologies (see Table~\ref{tab:sok-mapping}). This creates a link of what is potentially detectable, even in cases where prior studies did not specify the techniques and technologies they targeted (see \autoref{sec:data-labeling}). 

\begin{table}[htbp]
  \caption{Web Artifact Definitions}
  \label{tab:artifact-defs}
  \centering
  \scriptsize
  \begin{tabularx}{\columnwidth}{rX}
    \toprule
    \textbf{Web Artifact}    & \textbf{Definition} \\
    \midrule
    \textbf{Request Path}          & The URL path and query string of each HTTP request (inc. fragment associated with a requested URL).\\
    \textbf{Protocol Metadata}     & Network, transport, and application metadata such as HTTP headers, TLS handshake parameters, IP/port/source address, and DNS query/response fields. \\
    \textbf{Web Page}              & The HTML document (DOM) and inline JavaScript, inc. observable runtime behavior in the page context. \\
    \textbf{Resource}              & External subresources fetched by the page such as images, stylesheets, or scripts. \\
    \textbf{Storage \& Cache}               & Client-side persistence mechanisms such as cookies, localStorage, ETags, cache API/service worker caches, and HTTP cache state when used to maintain identifiers. \\
    \bottomrule
  \end{tabularx}
\end{table}

\noindent\textbf{Current approaches can be classified into real-time and offline modes of detection \trend{\textbf{T6}}.}
The mode of detection is determined by either the features and, consequently, the classifier design or by the authors choice, which then dictates feature design. 
For example, certain features cannot be computed in real-time on a single page load, such as aggregate features (in-degree of a host) or the centrality of a tracker~\cite{raschke_tex-graph_2023, siby-2022, adgraph}.    
Many studies describe an offline scenario by positioning their detector as an alternative or support to filter lists, e.g., automatically producing rules for later use or maintenance~\cite{siby-2022, bytecode-2023, rieder_beyond_2025}. 
Real-time detectors~\cite{adgraph, yang_wtagraph_2022} rely on immediately observable features that are available during page load. However, these detectors necessitate additional considerations with respect to feature computation time and memory usage. \textsc{AdFlush} or \textsc{Deep Tracking Detector} are one of the recent advances that reflect a maturation from purely academic offline classifiers to deployable real-time defenses~\cite{lee_adflush_2024, uroz-dl}. 

\noindent\textbf{Most research formulates detection as a supervised binary classification problem, paired with the use of traditional ML-models and curated features \trend{\textbf{T7}}.} 
The overall process, problem formulation, and learning family remained relatively constant over the years with few exceptions such as one-class~\cite{ikram}, unsupervised~\cite{metwalley, uroz-dl}, or federated learning~\cite{annamalai-2024}. 
Although detectors using supervised learning have been shown to yield good results, they may inherit systemic biases from filter lists (see \autoref{sec:data-labeling}) and risk learning spurious correlates instead of causal tracking behavior \limitation{\textbf{L4}}. 
For their performance, tree ensembles such as random forests or boosting algorithms are chosen, while deep learning models have been sparingly utilized~\cite{campeny-2024, uroz-dl}. 

\begin{table}[htbp]
  \caption{Feature Type Definitions}
  \label{tab:feature-defs}
  \centering
  \scriptsize
  \begin{tabularx}{\columnwidth}{rX}
    \toprule
    \textbf{Feature Type}    & \textbf{Definition} \\
    \midrule
    \textbf{Content}      & Depends only on lexical/byte strings present in $\mathcal{A}$. \\
    \textbf{Statistical}  & Computes order-invariant aggregates over observable quantities tied to $\mathcal{A}$. \\
    \textbf{Structural}   & Is a function of within-page graph/structure derived from $a_i\in \mathcal{A}$, encoding topology or hierarchy. \\
    \textbf{Flow}         & Summarizes directed information flows from sources to sinks induced by $a_i\in \mathcal{A}$ in the execution context. \\
    \textbf{Sequential}   & Encodes order- or time-dependent event sequences linked to $a_i\in \mathcal{A}$, where permutation changes the feature value. \\
    \textbf{Relational}   & Depends on external or cross-site context of an $a_i\in \mathcal{A}$ or its relation to the top-level site that are not derived from local payload or structure alone. \\ 
    \bottomrule
  \end{tabularx}
\end{table}

\begin{table*}[ht]
	\centering
        \scriptsize
        \caption{Mapping of Web Artifacts to Existing Web Tracking Techniques and Technologies} 
	\renewcommand{\arraystretch}{1.2}
          \begin{tabular}{r||r||*{22}{c}}
            \toprule
            \multicolumn{1}{r}{}
              & \multicolumn{1}{@{}r||}{} 
              & \multicolumn{13}{c}{\textbf{Stateful}}
              & \multicolumn{3}{c}{}
              & \multicolumn{3}{c}{\textbf{Stateless}} \\ 
            \cmidrule(lr){3-15} \cmidrule(lr){16-24}
            \multicolumn{1}{c}{}
                & \multicolumn{1}{@{}r||}{\rotatebox{90}{\shortstack{\textbf{Tracking Techniques}\\ \textbf{and Technologies}}}}
            & \rotatebox{90}{Cookies~\cite{Kristol2001Cookies}}
            & \rotatebox{90}{Cookie Syncing~\cite{englehardt}}
            & \rotatebox{90}{Supercookies~\cite{soltani}}
            & \rotatebox{90}{CNAME cloaking~\cite{dao-cname-journal}}
            & \rotatebox{90}{Pixels~\cite{web-bugs-early}\kern1pt/\kern1pt Tags~\cite{tags}}
            & \rotatebox{90}{ETag~\cite{ayenson}}
            & \rotatebox{90}{Favicon~\cite{favicons}}
            & \rotatebox{90}{UID Smuggling~\cite{randall}}
            & \rotatebox{90}{Bounce Tracking~\cite{randall}}
            & \rotatebox{90}{Redirect Link Tracking~\cite{koop2020}}
            & \rotatebox{90}{Link Decoration~\cite{takata}}
            & \rotatebox{90}{Session-replay Scripts~\cite{NoBoundariesSessionReplay}}
            & \rotatebox{90}{Local and Session storage~\cite{acar-measurement}}
            & \rotatebox{90}{Referer Header~\cite{krishnamurthy-referer}}
            & \rotatebox{90}{IP-address tracking~\cite{ip-address}}
            & \rotatebox{90}{Canvas FP~\cite{mowery}}
            & \rotatebox{90}{WebGL FP~\cite{webgl}}
            & \rotatebox{90}{AudioContext FP~\cite{englehardt}}
            & \rotatebox{90}{Browser Extension FP~\cite{browser-extension-fp}}
            & \rotatebox{90}{Font FP~\cite{fpdetective}}
            & \rotatebox{90}{Hardware-based FP~\cite{webgl}} 
            & \rotatebox{90}{DNS-based Tracking~\cite{dns-tracking}}\\ 
            \midrule
            \multirow{5}{*}{\rotatebox{90}{\textbf{Web Artifacts}}}
           & \cellcolor{gray!15} Request Path 
            & \cellcolor{gray!15} $\circ$ 
              & \cellcolor{gray!15} $\bullet$ 
              & \cellcolor{gray!15} $\circ$ 
              & \cellcolor{gray!15} $\bullet$ 
              & \cellcolor{gray!15} $\bullet$ 
              & \cellcolor{gray!15} $\circ$ 
              & \cellcolor{gray!15} $\bullet$ 
              & \cellcolor{gray!15} $\bullet$ 
              & \cellcolor{gray!15} $\bullet$ 
              & \cellcolor{gray!15} $\bullet$
              & \cellcolor{gray!15} $\bullet$
              & \cellcolor{gray!15} $\circ$ 
              & \cellcolor{gray!15} $\circ$ 
              & \cellcolor{gray!15} $\circ$ 
              & \cellcolor{gray!15} $\circ$ 
              & \cellcolor{gray!15} $\circ$ 
              & \cellcolor{gray!15} $\circ$ 
              & \cellcolor{gray!15} $\circ$ 
              & \cellcolor{gray!15} $\circ$ 
              & \cellcolor{gray!15} $\circ$ 
              & \cellcolor{gray!15} $\circ$
              & \cellcolor{gray!15} $\circ$ \\
            & \cellcolor{white} Protocol Metadata 
            & \cellcolor{white} $\bullet$ 
              & \cellcolor{white} $\circ$ 
              & \cellcolor{white} $\bullet$ 
              & \cellcolor{white} $\bullet$ 
              & \cellcolor{white} $\circ$ 
              & \cellcolor{white} $\bullet$ 
              & \cellcolor{white} $\circ$ 
              & \cellcolor{white} $\circ$ 
              & \cellcolor{white} $\circ$ 
              & \cellcolor{white} $\circ$
              & \cellcolor{white} $\circ$
              & \cellcolor{white} $\circ$ 
              & \cellcolor{white} $\circ$ 
              & \cellcolor{white} $\bullet$ 
              & \cellcolor{white} $\bullet$ 
              & \cellcolor{white} $\circ$ 
              & \cellcolor{white} $\circ$ 
              & \cellcolor{white} $\circ$ 
              & \cellcolor{white} $\circ$ 
              & \cellcolor{white} $\circ$ 
              & \cellcolor{white} $\circ$
              & \cellcolor{white} $\bullet$ \\
           & \cellcolor{gray!15} Web Page  
            & \cellcolor{gray!15} $\circ$ 
              & \cellcolor{gray!15} $\bullet$ 
              & \cellcolor{gray!15} $\circ$ 
              & \cellcolor{gray!15} $\circ$ 
              & \cellcolor{gray!15} $\bullet$ 
              & \cellcolor{gray!15} $\circ$ 
              & \cellcolor{gray!15} $\bullet$ 
              & \cellcolor{gray!15} $\bullet$ 
              & \cellcolor{gray!15} $\circ$ 
              & \cellcolor{gray!15} $\circ$
              & \cellcolor{gray!15} $\bullet$ 
              & \cellcolor{gray!15} $\bullet$ 
              & \cellcolor{gray!15} $\bullet$ 
              & \cellcolor{gray!15} $\circ$ 
              & \cellcolor{gray!15} $\circ$ 
              & \cellcolor{gray!15} $\bullet$ 
              & \cellcolor{gray!15} $\bullet$ 
              & \cellcolor{gray!15} $\bullet$ 
              & \cellcolor{gray!15} $\bullet$ 
              & \cellcolor{gray!15} $\bullet$ 
              & \cellcolor{gray!15} $\bullet$
              & \cellcolor{gray!15} $\circ$ \\
            & \cellcolor{white} Resource  
            & \cellcolor{white} $\circ$ 
              & \cellcolor{white} $\bullet$ 
              & \cellcolor{white} $\circ$ 
              & \cellcolor{white} $\circ$ 
              & \cellcolor{white} $\bullet$ 
              & \cellcolor{white} $\circ$ 
              & \cellcolor{white} $\circ$ 
              & \cellcolor{white} $\circ$ 
              & \cellcolor{white} $\circ$ 
              & \cellcolor{white} $\circ$ 
              & \cellcolor{white} $\circ$ 
              & \cellcolor{white} $\bullet$ 
              & \cellcolor{white} $\circ$ 
              & \cellcolor{white} $\circ$ 
              & \cellcolor{white} $\circ$ 
              & \cellcolor{white} $\bullet$ 
              & \cellcolor{white} $\bullet$ 
              & \cellcolor{white} $\bullet$ 
              & \cellcolor{white} $\bullet$ 
              & \cellcolor{white} $\bullet$ 
              & \cellcolor{white} $\bullet$
              & \cellcolor{white} $\circ$ \\
           & \cellcolor{gray!15} Storage \& Cache 
           & \cellcolor{gray!15} $\bullet$ 
              & \cellcolor{gray!15} $\circ$ 
              & \cellcolor{gray!15} $\bullet$ 
              & \cellcolor{gray!15} $\circ$ 
              & \cellcolor{gray!15} $\circ$ 
              & \cellcolor{gray!15} $\bullet$ 
              & \cellcolor{gray!15} $\circ$ 
              & \cellcolor{gray!15} $\circ$ 
              & \cellcolor{gray!15} $\circ$
              & \cellcolor{gray!15} $\circ$
              & \cellcolor{gray!15} $\circ$ 
              & \cellcolor{gray!15} $\circ$ 
              & \cellcolor{gray!15} $\bullet$ 
              & \cellcolor{gray!15} $\circ$ 
              & \cellcolor{gray!15} $\circ$ 
              & \cellcolor{gray!15} $\circ$ 
              & \cellcolor{gray!15} $\circ$ 
              & \cellcolor{gray!15} $\circ$ 
              & \cellcolor{gray!15} $\circ$ 
              & \cellcolor{gray!15} $\circ$ 
              & \cellcolor{gray!15} $\circ$
              & \cellcolor{gray!15} $\circ$ \\
            \bottomrule
          \end{tabular}
        \begin{minipage}{\textwidth}
        \vspace{1mm}
        \centering
        \scriptsize
        Mapping of the web artifacts to the potential detection of tracking techniques and technologies. A $\bullet$ denotes that the web artifact (a classifier target for a given target) may identify the corresponding tracking technique and technology, while a $\circ$ denotes that it cannot. A detector can also combine multiple web artifacts to increase coverage. 
        \end{minipage}
 \label{tab:sok-mapping}
\end{table*}

\noindent\textbf{A large body of features has been developed, yet their subsequent use in research remains rare \opportunity{\textbf{O8}}.} 
We categorize features into six types (Table~\ref{tab:feature-defs}, $\mathcal{A}$ as crawled records). Depending on the web artifact(s), authors employ one or more types. For instance, a pure HTTP header-based detector uses only content features while a detector based on abstract syntax trees (ASTs) utilizes multiple types. 
Researchers usually curate (candidate) features using their domain expertise and knowledge
Feature selection, pruning, or optimization can then be automated using, e.g., information gain~\cite{iqbal-fingerprinting-2021}, $\chi^2$ algorithm~\cite{wu2016}, recursive feature elimination~\cite{trackadvisor, lee_adflush_2024}, simple heuristics~\cite{rieder_beyond_2025}, or dimensionality reduction methods~\cite{lee_adflush_2024}. To address the challenges of scale, complexity, and high-dimensionality of web artifacts, automated feature extraction methods were used by deep learning and NLP-based detection studies~\cite{uroz-dl, campeny-2024, qiang_deepfpd_2024, ast-trans}. A handful of studies utilized past feature sets, with \textsc{AdFlush}~\cite{lee_adflush_2024} representing a comprehensive inclusion analysis to identify an optimized set (based on~\cite{adgraph, siby-2022, yang_wtagraph_2022}). Others follow previous feature extraction methods~\cite{shuang_dumviri_2025} (based on~\cite{adgraph, siby-2022}); incorporate findings~\cite{guarino} (based on~\cite{trackadvisor, gugelmann}); augment features from ~\cite{adgraph} in~\cite{siby-2022}; or acknowledge and confirm~\cite{rieder_beyond_2025}. 

\noindent\textbf{Detectors are increasingly using graph-based approaches that target the structural properties of trackers \trend{\textbf{T8}}.} 
The selection of feature types is typically a trade-off between computational complexity, cost, and robustness. Content and statistical features can be more easily evaded (entailing a performance drop) while structural and flow features (enabled by graphs) are harder to obfuscate. However, the risk for data poisoning or manipulation still exists (see \autoref{sec:performance-evaluation}). At the core of graph-based approaches are ASTs, which capture the semantic meaning of JavaScript files and especially API keywords in a usable and traversable format for feature extraction. ASTs were first used by Orr et al.~\cite{orr} in 2012. ASTs can either be used alone~\cite{trackerdetector, castell-uroz_astrack_2023} or in tandem with other methods, e.g., to construct program dependency graphs~\cite{ikram} or with other graph structures~\cite{lee_adcpg_2023}, called a Code Property Graph which includes a Control Flow Graph and Program Dependency graph~\cite{cpg}. Approaches vary in their definition of nodes (e.g., hosts~\cite{kalavri}, resources~\cite{cookiegraph-2023}) and edges (e.g., HTTP requests~\cite{raschke_tex-graph_2023}, and interaction types~\cite{kargaran-2021}). These variations extend to the scope on what they actually capture, e.g., code structure or semantics (static)~\cite{bahrami_fp-radar_2022, qiang_deepfpd_2024}, runtime interactions or information flow during a page load (dynamic)~\cite{siby-2022, amjad-2024}, or macroscopic relationships with aggregated data across websites (network and host dependency graphs)~\cite{raschke_tex-graph_2023, kalavri}. 
Graph-based approaches have been demonstrated to facilitate the detection of trackers and tracking at a more fine-granular level. These approaches also enable the detection of mixed resources at the script or function level as well~\cite{amjad-2024, castell-uroz_astrack_2023}. 

\noindent\textbf{Towards explainable cross-domain and mixture-of-experts detectors with multi-label outputs in a continual learning setting \opportunity{\textbf{O9}}.} 
Today, researchers are developing individual detectors using either similar or vastly different methodologies. Some approaches combine multiple web artifacts while others focus on one. Future research could explore mixture-of-experts models that combine past approaches. 
Learning should be inductive and continual (inc. federated learning\cite{annamalai-2024}) and explicitly adversarial-aware (counterfactual and evasion tests, inc. API substitution, packing, and CNAME aliasing \cite{dao}), rather than relying solely on robust features. Although graph neural networks and transformers over AST sequences have improved accuracy \cite{yang_wtagraph_2022,ast-trans,campeny-2024}, hybrid graph-transformer designs and foundation models for detection should be pursued, with compression or distillation as first-class requirements for deployment. Finally, detectors should emit policy-aligned, multi-label outputs (e.g., analytics, ads, or fingerprinting) with built-in explainability to enable precise remediation and curation. They should also be evaluated under inductive (domain/site hold-out, temporal shifts) and transductive (graph-aware masks) regimes to demonstrate robustness.

\subsection{Performance Evaluation}\label{sec:performance-evaluation}
The penultimate step in the pipeline is the performance evaluation using various metrics, in- and out-of-distribution datasets, experiments, and classifier optimizations. While the field is shifting towards more comprehensive evaluations, there is room for improvement in how we test detectors in a dynamic and adversarial environments. 

\noindent\textbf{Prior work focuses on standard threshold-dependent performance metrics to assess detectors in offline lab evaluations using static data from a single point in time \limitation{\textbf{L5}}.} 
The majority of the literature, particularly studies before 2020, follows the procedure of training a new detector by: selecting a set of models; using k-fold cross-validation or hold-out splits on one dataset; and evaluating the trained classifiers using Precision, Recall, Accuracy, and F1-score (inc. other confusion matrix metrics). 
Most detectors are evaluated on offline crawled datasets in a controlled lab-environment rather than in diverse, live user (real-world) environments. This distances the results from operational reality, as deployed detectors and user studies are uncommon. Additionally, if the study did not use a stratified sampling approach during the crawling process, popularity bias is inherited in the evaluation. The results are then based on domain prevalence and the performance metrics can be distorted~\cite{kargaran-2021}. 
Although the described procedure is not inherently flawed, it prioritizes achieving high scores (in comparison to filter lists) while neglecting the dynamic, adversarial, and resource-constrained (client-side) nature of the web. Recent studies have shown that future research should focus on designing fundamentally more robust, efficient, and user-centric detectors~\cite{lee_adflush_2024, siby-2022} \opportunity{\textbf{O10}}.

\noindent\textbf{The current lack of baselines, inappropriate performance metrics, real-world testing, and diverse evaluation data can misrepresent the actual performance of detectors and hinder comparability \limitation{\textbf{L6}}.} 
The evaluation of a new detector is one way to state its efficacy and added value to the state-of-the-art. However, we observe several limitations that may overstate the actual efficacy while hindering fair comparisons. Baselines are rarely defined (simple and complex models), which limits the interpretation of the actual improvements -- one exception being filter lists that are commonly used for labeling and as a baseline. This is further exacerbated by the lack of publicly available and functioning artifacts (\autoref{sec:rep}). 
Studies focus on using threshold-dependent performance metrics, which only provide a static, single snapshot of a detector's performance. This creates an illusion of sufficiency where high scores are treated as the definitive measure. Longitudinal experiments further highlight measurable drifts, underscoring that single snapshot tests are misrepresenting~\cite{dao, khaleesi-2022}. 
Threshold-independent performance metrics summarize a detector's overall efficacy and provide a more realistic assessment~\recommendation{\textbf{R7}}. Furthermore, selected performance metrics do not always account for the true data distribution, which is evidenced to be imbalanced~\cite{rieder_beyond_2025, englehardt}. This class imbalance is also endemic to specific techniques and technologies, such as fingerprinting~\cite{englehardt, panopticlick, iqbal-fingerprinting-2021}. In turn, this results in a misinterpretation of the base rate and an overestimation of performance. 

\noindent\textbf{Today, researchers conduct more experiments and analyses to evaluate detectors with a recent focus on generalizability, robustness, and breakage \trend{\textbf{T9}}.} 
Research has shifted from assessing the core performance and filter list comparisons (inc. the reporting of unseen trackers) to experiments that consider a detector in a live environment. 
We distinguish between: generalizability (unseen data without adversarial manipulation); robustness (performance stability under perturbations, inc. adversarial manipulations); and breakage (loss of legitimate website functionality). Generalizability focuses primarily on three types: cross-validation; longitudinal; and cross- or distributed-environments. 
To assess breakage caused by a detector's misclassifications or mixed-trackers, researchers conduct: manual assessment with human reviewers~\cite{iqbal-fingerprinting-2021, shuang_dumviri_2025}; or automated assessments such as computer vision~\cite{castell-uroz_astrack_2023}; breakage metrics based on reloading behavior~\cite{yu}; or a specialized breakage detector based on differential features~\cite{shuang_dumviri_2025} (results could also translate from~\cite{sinbad, smith}; or from anti-adblock detection~\cite{anti-adblock}). 
Robustness experiments address evasion and circumvention techniques, accompanied by analysis and discussion on feature stability. We provide a first classification in Table~\ref{table:ev-cv-techniques}.  
All three foci determine a detector's performance in a live environment and a user's acceptance rate in terms of usefulness and usability. However, robustness and especially breakage, require deployable detectors to be tested empirically. Most studies lack these tests (\autoref{tab:tracker-detector-table}), thus the majority never load live websites with their detector and provide no empirical evidence that blocking decisions preserve site functionality \limitation{\textbf{L7}}. 

\definecolor{amber}{rgb}{1.0, 0.49, 0.0}
\definecolor{crimson}{rgb}{0.86, 0.08, 0.24}
\newcolumntype{L}[1]{>{\raggedleft\arraybackslash}p{#1}}
\newcolumntype{P}[1]{>{\centering\arraybackslash}p{#1}}

\begin{table}[ht]
    \centering
    \caption{Evasion and Circumvention Techniques.}
    \ssmall
	\begin{tabularx}{\columnwidth}{L{0.3\linewidth}X}
            \toprule
		\textbf{Category} & \textbf{Technique}
            \\
		\midrule
            \cellcolor{amber!20} \textbf{URL and Endpoint Manipulation}
            & Replacing tracking-related tokens, URL length padding, Rotating second-level domains and subdomains for tracking endpoints
            \\ 
            \cellcolor{amber!20} \textbf{Identifier Parameter Manipulation}
            & Renaming, splitting, combining, or encoding link-decoration parameters  
            \\ 
            \cellcolor{amber!20} \textbf{State Namespace Randomization} 
            & Cookie chunking, Randomizing names of identifier-carrying cookies or keys
            \\ 
            \cellcolor{amber!20} \textbf{Protocol Manipulation} 
            & Toggling/removing headers, request-chain shortening 
            \\ 
            \cellcolor{amber!20} \textbf{Code Transformation}
            & HTML inlining, Bundling tracking with benign functionality, code obfuscation/transformations, minification 
            \\    
            \cellcolor{amber!20} \textbf{Distributed Behavior} 
            & Distributing functionality across colluding code units 
            \\ 
            \cellcolor{crimson!20} \textbf{Identity Masking}
            &  CNAME cloaking, Server-side tracking migration
            \\ 
		\bottomrule
	\end{tabularx}
    \begin{minipage}{\columnwidth}
    \vspace{1mm}
    \centering
    \scriptsize
    \textbf{Legend:} Techniques can be either \textcolor{amber}{evasive} or \textcolor{crimson}{circumventive} Some techniques and technologies could be classified as evasion or circumvention but were excluded. 
    \end{minipage}
 \label{table:ev-cv-techniques}
\end{table}

\noindent\textbf{The community would benefit from a standardized evaluation framework, an adversarial simulation testbed, and a public benchmark dataset to enable reproducible and valid comparisons between detectors to advance the state-of-the-art \opportunity{\textbf{O11}}.} 
Current evaluation methodologies in the field are heterogeneous, which hinders direct comparisons between detectors. This is exacerbated by the absence of a benchmark dataset (inc. longditudinal and cross-browser data) (\autoref{sec:data-collection}). Metric-based comparisons alone are therefore unreliable and do not reflect the actual performance gains. 
Although testing evasion and circumvention has become more common in recent years, a study on adversarial attacks (inc. a testbed) towards detectors in a live environment is still missing. 
A testbed should include a diverse and evolving set of attacks (inc. evasion and circumvention) while being maintained by the community. 
It would contribute to a standardized evaluation methodology for live environments (or at least approximating the real-world conditions) that does not exist yet. 
This could lay the ground for future work on adversarial ML, such as attack surfaces, adversarial training, robust optimization, and poisoning. 
Future research does not lie in designing marginally more accurate detectors on static, single snapshot datasets, but in designing fundamentally more resilient and efficient ones across various experiments, diverse datasets, and environments.

\subsection{Deployment and Operations}\label{sec:deployandoperate}
One core motivation of web tracker detection is to preserve user privacy. 
This requires detectors to be deployed either: client-side~\cite{lee_adflush_2024, yang_wtagraph_2022, cozza}; hybrid/distributed (detector split across multiple entities)~\cite{annamalai-2024}; or network-level~\cite{lee_net-track_2023, dao-cname-journal}.
Detectors can then be operated either offline (run on logs or crawls, e.g., filter list generation) or real-time (can directly intervene, e.g., live detection). 
Each scenario has its own number of requirements to ensure its expected functionality in various use-cases and operational modes. Especially real-time detectors are subject to low latency and high throughput while working in a resource-constrained environment in terms of memory, computational power, and model size. These requirements should be translated to relevant deployment metrics (e.g., from \cite{cozza, fpflow-2022, lee_adflush_2024}, \cite{shuba}$^\ddagger$) and added to the aforementioned standardized evaluation framework and testbed. 

\noindent\textbf{Researchers are increasingly exploring and testing their detectors in client-side deployment scenarios \trend{\textbf{T10}}.} 
Although three early works around the 2000s already addressed deployment~\cite{kushmerick}\textsuperscript{$\ddag$}, \cite{alsaidDetectingWebBugs2003}\textsuperscript{$\ddag$}, \cite{fonseca}\textsuperscript{$\ddag$}, the topic regained traction only after 2015, and more after 2020. This can be traced back to parallel developments on robustness, breakage, and generalizability. We classify detectors according to three deployment readiness levels (DLR): (DLR-1) detector is offline and deployment is only discussed~\cite{meyer_detecting_2024, castell-uroz-tracksign_2021}; (DLR-2) detector is a prototype and evaluated in a lab-environment (inc. operational performance metrics)~\cite{lee_net-track_2023}; (DLR-3) detector is adjusted for deployment and tested in a real-world environment or under real-world settings~\cite{lee_adflush_2024}. The operational mode, deployment and application scenario should be considered early on in the research process~\recommendation{\textbf{R8}}. 

\noindent\textbf{Deployment allows for more realistic assessments and opens new, currently under-researched areas in detection \opportunity{\textbf{O12}}.} 
A small subset of detectors have been implemented as: browser extensions~\cite{lee_adflush_2024, polcak_jshelter_2023, khaleesi-2022}; modified browsers~\cite{fpflow-2022, wu2016}; or as part of a proxy~\cite{ast-trans}. These are typically evaluated in limited experiments rather than long-running, user-facing deployments. Real-world behavior such as: concept drift under continuous use; user-visible breakage rates; latency and resource overhead on a set of heterogeneous devices; effectiveness of different classifier approaches in everyday browsing; and adversarial adaptation by trackers remain largely unexplored. Deployment would therefore not only improve the evaluation, but also create a new empirical basis for studying online ML and federated learning; adversarial robustness; governance model questions; and user feedback (inc. human-in-the-loop; user vs detector decisions; remediation strategies to restore functionality without degrading privacy protection).

\section{Reproducibility of Web Tracker Detectors}\label{sec:rep}
We review each paper for references to publicly available resources and list them in Table~\ref{tab:tracker-detector-table}. Each paper can produce at least one artifact: the code; and produce or report the use of existing artifacts: the dataset(s); models; and ground-truths. The goal is to analyze the state of the artifacts and to reproduce a set of artifacts to determine whether a full reproducibility study is needed. 

\noindent\textbf{Results of Reproducibility Experiments}
Of all the papers reviewed, most did not provide complete artifacts to reproduce their results. We only conducted experiments where the trained model or the dataset was provided and the code was functional.
In all seven cases, we were able to reproduce similar values compared to stated results in the studies (Table~\ref{tab:repro-results}) -- although some minor troubleshooting was necessary. \textsc{Duumviri}'s, \textsc{PURL}'s, and \textsc{AdFlush}'s artifact were reproduced without any issues and are a good example for future studies \recommendation{\textbf{R9}}. However, we were not able to reproduce all papers with artifacts due to issues. 
We emphasize that some of the excluded studies in Table~\ref{tab:repro-results} were already reproduced by other studies, even if we did not manage to do so (one did not work halfway)~\cite{khaleesi-2022, siby-2022, adgraph, iqbal-fingerprinting-2021},~\cite{bhagavatula}$^\ddagger$. These studies should be treated with the same level of reproducibility. 
Future work should conduct a full reproducibility or replicability study with publicly available (containerized) artifacts \opportunity{\textbf{O13}}. We postulate that this could confirm and strengthen our trust in previous research results and enable similar benchmark studies that were conducted during the 2010s for filter lists and tracker defense tools~\cite{mazel, ruffell, snyder, wills, traverso, merzdovnik, alrizah}. 

\begin{table}[ht]
	\centering
        \scriptsize
        \caption{Results of Reproducibility Experiments}
        \begin{tabular*}{\columnwidth}{@{\extracolsep{\fill}} rlccc}
        \toprule
		\textbf{Ref} & \textbf{Model} & \textbf{Accuracy} & \textbf{F1-Score} & \textbf{A-Ref}\\
		\midrule
            \textsc{AdFlush}\cite{lee_adflush_2024} & 
            GBM & 
            0.9865 & 
            0.9829 & 
            \cite{chaejin-lim-2024-10682483}
            \\
             & 
             & 
            (0.9800) & 
            (0.9800) & 
            \\
            \textsc{Duumviri}\cite{shuang_dumviri_2025} & 
            XGB & 
            0.9661 & 
            0.0000$^\dagger$ & 
            \cite{duumviri-git} 
            \\ %
            & 
             & 
            (0.9385) & 
            (0.8595) & 
            \\
            \textsc{PURL}\cite{munir_purl_2024} & 
            RF & 
            0.991 & 
            Not reported &
            \cite{purl-git, purl-dataset}
            \\ %
             & 
             & 
            (0.9874) & 
            Not reported & 
            \\
            \textsc{WTAGRAPH}\cite{yang_wtagraph_2022} & 
            GNN & 
            0.9840 & 
            0.9796 & 
            \cite{wtagraph-git, wtagraph-dataset}
            \\ %
            & 
            & 
            (0.9790) & 
            (0.9730) & 
            \\
            \textsc{Response}\cite{rieder_beyond_2025} & 
            ET & 
            0.958 & 
            0.902 & 
            \cite{zenodo-dataset, zenodo-new-dataset},
            \\ %
            & 
            & 
            (0.9600) & 
            (0.9310) & \cite{response-git} 
            \\
            \textsc{T.EX Graph}\cite{raschke_tex-graph_2023} & 
            XBG & 
            0.8750 & 
            0.8400 & 
            \cite{t.ex-git, zenodo-dataset}
            \\ %
            & 
            & 
            (0.8830) & 
            (0.8670) & 
            \\
            \textsc{Bytecode}\cite{bytecode-2023} & 
            RF & 
            0.9645 & 
            0.9589 & 
            \cite{byte-learn-git, byte-learn-dataset} 
            \\ %
            & 
            & 
            (0.9708) & 
            (0.9556) & 
            \\
		\bottomrule
        \end{tabular*}
        \begin{minipage}{\columnwidth}
        \vspace{1mm}
        \centering
        \scriptsize
        We reran the original experiments that the authors from the reproduced studies defined (experiment: detector performance with the in-distribution test set). The reported results are from our experiments while the original results from the studies are given in parentheses. $^\dagger$published code did not export the metric. A-REF stands for artifact reference (code and/or dataset(s)). 
        \end{minipage}
 \label{tab:repro-results}
\end{table}

\noindent\textbf{Artifact Issues and Recommendations}
Although the experiments were not conducted for all papers, we want to emphasize that this does not dismiss of a study's findings or methodology. Considering the accumulated evidence of detection performance over the last 25 years, we can confidently conclude that detectors work. Our assessment of the quality of evidence and the experiments is only a preliminary attempt to understand the current state.
Published and well-maintained artifacts facilitate future research, e.g., by enabling baselines to assess the improvements. We observe an increase in published artifacts over time and experiments showed statistically similar results compared to the stated values in the corresponding studies. 
We identify the following artifact issues and recommendations for future research: 

\noindent\textbf{Lack of Availability.} From all identified papers, only a minority provided dataset(s) and code. All artifacts (inc. model export) should be published using a long-term hosting solution with versioning, such as Zenodo\cite{Zenodo}. 

\noindent\textbf{Lack of Documentation.} Some artifacts lack concrete instructions to build and execute the environment and experiments. Information regarding potential problems, hotfixes, or time and resource requirements to help a third-party are rarely reported. At least one file on how to use the artifacts should be included. Examples are  USENIX\cite{USENIXSec25Appendix} and PETS\cite{PoPETsArtifactEvaluation}. Moreover, the concrete version number of the used ground-truths, in case of \textsc{EasyList} or \textsc{EasyPrivacy} the commit hash, should be documented as these filter lists change over time. If versions are not reported, researchers cannot determine whether differences in results stem from the detector or from changes in the underlying ground truth. 

\noindent\textbf{Dependency Conflicts.} 
Dependencies are either: without version numbers; obsolete; or outdated dependencies. These issues lead to dependency conflicts that are difficult to solve. Lastly, environment manifest files were usually missing as well.
Including dependency manifest files ensures the correct environmental setup. Dependency management tools can further help. To avoid potential conflicts while improving maintainability and performance, dependencies should be reduced to a minimum. 

\noindent\textbf{Missing Containerization.} Related to the last issue is the lack of containerized artifacts to encapsulate dependencies, code, and runtime configurations into one, immutable environment package. Containerization reduces reproducibility issues, minimizes inconsistencies across systems, and ensures portable, and verifiable artifacts. 

\noindent\textbf{Undisclosed Methodology Steps.} The pre-processing stage in the ML workflow can be shallow or missing. Without knowing the concrete steps, it can be challenging to reproduce or replicate the findings. Documenting each step either in the paper or artifact. Other research scientific disciplines require electronic laboratory notebooks to document the research work from planning to evaluation and to ensure good scientific practice~\cite{gerlach-2020, higgins-2022} -- a method that could be adapted in this field. 

\noindent\textbf{Resource and Time Constraints.} A third-party's ability to use published artifacts can be limited by substantial computational requirements. Optimizing artifacts to ensure their usefulness for a broader audience (especially smaller research groups with limited resources), and thus their applicability in future studies. 

Although, the issues and recommendations discussed above are common reproducibility challenges in computational research. They are particularly consequential in the web tracker detection domain. Detector performance depends on, e.g., volatile web data; browser and crawler configurations; and evolving labeling sources. As a result, insufficiently documented or unavailable artifacts limit reuse, verification, comparison and longitudinal interpretation of reported results. 

\section{Discussion}\label{sec:discussion} 
Following our synthesis on detectors, we now focus on broader topics that did not fit in the previous sections, and how to situate our work ethically. 

\noindent\textbf{Ethical Considerations.}
We conduct an ethical analysis of our SoK paper for potential ethical problems in our research (\autoref{sec:ethics}). Using the deontological and consequentialist frameworks, we conclude that the potential benefits of our research outweigh the risks and that our actions adhered to ethical principles. 
However, the absence of an ethical screening criterion is a shortcoming in our eligibility assessment (\autoref{sec:sys-review}). Our SoK paper relies and integrates previous work, and the ethics of these studies extend to our work. 

Authors usually recognize the effects detection has on relevant stakeholders (see Fig.~\ref{fig:stakeholders}). At the same time, no study conducted or reported an ethical analysis of their work, although detection raises societal and technical concerns. 
We encourage authors to assess the ethics of their work, e.g., a detector has dual-use potential: being a privacy-enhancing tool for users; and as instruments to inform tracker provider's evasion and circumvention strategies. Furthermore, false positives can harm benign website providers and thus their economic survival. These few examples demonstrate the need to reflect the potential benefits and harms. We refer to the Menlo Report on ``Ethical Principle Guiding Information and Communication Technology Research'' as a starting point~\cite{menlo_report_2012}. 

\noindent\textbf{Adoption Barriers and Governance.}
Detectors are not isolated entities and subject to potential technical, economic, and socio-legal barriers. Widespread adoption could threaten ad-based revenue models. Publishers, ad- and tracker providers, and some browser vendors may have weak incentives to integrate detectors, unless users demand so. User adoption depends on usability and decision support. Without careful human-computer-interaction design, detectors risk decision fatigue (for mixed trackers and breakage); misconfigurations; and low trust. Researchers could explore nudging, consent, and explainable AI for these points. 

Our systematization suggests that the limited real-world adoption of detectors may be due to a mismatch between research optimization targets, and deployment requirements. In practice, manually curated filter lists have proven to be operationally mature. They are simple to deploy, computationally lightweight, interpretable, and incrementally maintainable. Furthermore, they are already embedded in browsers and extensions. Although recent work considers deployment and client-side operation (see~\autoref{sec:deployandoperate}), only a limited number of detectors have been shown to reach a deployment-readiness level that corresponds to filter lists. 
However, enhancing detection performance is inadequate for ensuring broader adoption. Moving forward, the community should prioritize deployment-oriented evaluations under real-world settings, computationally inexpensive designs, interpretable and user-actionable decisions, and maintainable detectors. A viable approach to these challenges may be hybrid approaches~\cite{khaleesi-2022} that combine filter lists as a low-cost first layer, complemented by the adaptability of learning-based detectors to support rule generation or second layer decisions for cases that are ambiguous or previously unseen. 

Questions of governance arise around legal status and data ownership, e.g., detector providers must collect and process web telemetry (maybe user data) to train and update detectors in a federated learning setting. This raises compliance questions under data protection law. Different business and governance models, such as: open-source; community-driven projects; browser-integrated protections; enterprise; or paid services, imply different accountability structures. User-favorable models may be hard to align with commercial incentives and stakeholder interests. Lastly, detectors may benefit users of certain browsers, regions, or technical literacy levels (blocking decisions must be explainable rather than being binary values). The benefit over existing protection and defense tools must be clear to users. 

\noindent\textbf{Constraints on Detection.} Detection research is shaped by the tracking mechanisms it targets, the defenses and protections surrounding them, and the deployment role it is expected to fulfill. Tracking determines what can be observed and classified: different tracking families exist through different web artifacts and at different granularities (see Table~\ref{tab:sok-mapping}). These variations translate into a number of detector designs. Defense and protection then determine how detectors operate in practice. Real-time browser-side detectors must rely on immediately available signals under latency, memory, and breakage constraints. Offline detectors for auditing, measurement, or filter list generation can use aggregate or cross-site evidence that is unavailable during a live page load. 
Defense and protection also reshape the target itself. Browser-side measures such as storage partitioning, cookie restrictions, or query parameter removal have side-effects: limiting or suppressing some features; shifting towards first-party tracking; or creating incentives for evasion and circumvention~\cite{sok-web-tracking}. As a result, studies evaluate detectors against tracking mechanisms that operate under particular browser, defense, and deployment assumptions. This is why detector design, methodical choices, and claims of real-world relevance are only interpretable relative to these assumptions and tracking mechanisms. 

\noindent\textbf{Detection beyond the Web.}
Detection does not have to be constrained to the domain ``web'' and instead be applicable to other domains as well (e.g., using transfer learning). Existing work already showed differences between tracking behaviors on mobile versus desktop~\cite{yang}. Furthermore, in-app browsers and WebViews are also subject to tracking~\cite{webviews}, which require detectors to navigate in different execution environments -- and across devices. This could include devices for virtual and augmented reality, with potential tracking in WebXR environments~\cite{vr, webxr, son}. Today's detectors detect client-side tracking and with the increasing switch to server-side tracking on publisher's websites~\cite{sst-meta, fouad-server}, there is a pressing need for viable solutions. 
Finally, the recent rise of LLM-based agents and AI browsers (inc. browsing tools) introduce new surfaces for potential privacy and security vulnerabilities~\cite{shapira2025mindwebsecurityweb, ukani, zhang2025agent} -- while users may disclose more sensitive data than before~\cite{fair-game}. It is not clear how LLM-based agents interact with trackers and how detectors could enhance them. LLM-based agents could also be used: to simulate realistic, human-like interactions to enhance data collection; or to make privacy-preserving navigation on behalf of users. 

\section{Concluding Remarks}\label{sec:con}
Our SoK paper reviewed and systematized 59 primary papers and 16 supplementary studies to assess the current state of detection research and to identify its future path. We approached this topic with multiple perspectives: (i) stakeholder analysis; (ii) systematic review; (iii) systematization; (iv) taxonomy; (v) quality assessment of the evaluation; (vi) analysis of the state of artifacts; (vii) limited reproducibility experiments; and (viii) an ethical analysis of our research. 

For many years, researchers have been able to detect trackers with high accuracy. The core problem of tracker detection has been solved. Besides improving detectors, we argue that the field should shift its focus to the entire detector pipeline, particularly robust deployments. This includes conducting real-world tests because controlled lab tests do not represent the dynamic and adversarial nature of the web (and tracking). Shared benchmarks, clearer labeling definitions, realistic adversarial and breakage evaluations, and reproducible artifacts are required for this shift as well. 
Detectors and studies have reached a level of maturity, where users should be included -- despite the ethical, legal, technical, and recruiting challenges. Furthermore, detection research should evolve to assess user tracking besides browser-based client-side tracking such as server-side tracking and emerging agentic browsing environments.

\section*{Acknowledgments}
We thank our anonymous reviewers and our shepard for their valuable feedback. We thank from the Technische Universität Berlin, Yevheniia Strilets for her support on the ethical considerations and our student IoSL project group for their technical support in conducting the reproducibility experiments. 
This work was conducted at the SNET research group, which is part of T-Labs -- a public-private partnership between Technische Universität Berlin and Deutsche Telekom AG. 
\section*{Ethics Considerations}\label{sec:ethics}
Our SoK paper reviewed and systematized existing web tracker detection studies since the 1990s. While related work may not have discussed significant ethical issues, we acknowledge that it is insufficient to apply the analyses from past publications to our research. Therefore, we conduct an ethical evaluation in light of current technologies, societal expectations, and data protection requirements using the framework provided by the Menlo Report on ``Ethical Principle Guiding Information and Communication Technology Research''~\cite{menlo_report_2012}. 

\noindent\textbf{Stakeholders} We identify the following primary stakeholders of our research: 

\begin{enumerate}[itemsep=0.3mm, parsep=0pt]
    \item \textbf{Authors} of the reviewed studies, whose work or artifacts were analyzed and whose reputation or research trajectory could be influenced by our interpretation. 
    \item \textbf{Individuals} whose personal data may be contained in datasets used in the reviewed studies, even if anonymized, as their privacy could be indirectly affected by how such data is represented. 
    \item The \textbf{academic research community}, who may rely on our systematization of the current literature for future work. 
    \item The \textbf{general public}, given the broader societal implications of web tracking and privacy research. 
\end{enumerate}

\noindent\textbf{Potential Harms.} Considering the interests of all primary stakeholders, we recognize several potential harms, both tangible and intangible, that may occur in the short and long term as a result of our paper and its publication. Firstly, our research and experiments were conducted to the benefit of our community and do not pose any harm to these authors, nor with any intention to discredit their work (\textit{Respect for Persons}). However, there remains a possibility of reputational damage if our analysis and  interpretations are misrepresented or taken out-of-context. Secondly, although we treated all identified primary and supplementary studies impartially without bias toward any population, group, or person (\textit{Justice}), our eligibility criteria did not include an assessment of the ethical conduct of each study. Thus, it is possible that some included work did not meet current ethical standards. This risk is compounded by the fact that some studies are older and may have been conducted under ethical norms and data protection regulations that differ from current norms. Therefore, it is necessary to recognize the possibility that our paper indirectly legitimizes prior unethical research if it took place. While we trust our colleagues adhere to ethical research practices (and to the best of our knowledge, we have no reason to believe otherwise), this paper, nonetheless carries with the obligation to critically consider the ethical integrity of the studies it engages with. Finally, our systematization of detection approaches could be misused by adversaries to evade detection. 

\noindent\textbf{Mitigations and Unmitigated Risks.} To minimize these risks, we confined our systematization to publicly available sources. We did not recruit or interact with human participants, nor did we gather identifiable information about individuals. Experiments were only ran on code and data that was publicly released by the authors for others to use and we cited all sources. We did not disclose any implementation details beyond what authors already released. We also plan to inform authors post hoc and remain available to clarify our interpretations to reduce potential misrepresentations. Notwithstanding these measures, residual risks remain, such as the absence of an explicit ethics-screening criterion (and analysis), the inability to verify the compliance of older studies with current ethical and legal standards, and the potential for adversarial adaptation of tracking techniques. We acknowledge these risks and intent to address them in future work by developing explicit ethics assessment criteria and guidelines. 

\noindent\textbf{Ethical Analysis} From a consequentialist perspective, we have to weigh the anticipated benefits and harms against one another. The principal benefit is a holistic understanding of detection, which can guide future research and highlight new directions for advancing privacy-enhancing technologies. We argue that these contributions to the academic community, and the broader societal value of systematized privacy knowledge, outweigh the probable risks identified above. We nevertheless acknowledge that the full range of potential consequences cannot be entirely foreseen. 

From a deontological perspective, we must assess whether our actions adhered to the principles of \textit{Beneficence}, \textit{Respect for Persons}, \textit{Justice}, and \textit{Respect for Law and Public Interest}. Our study did not involve human subjects and posed no direct violations of rights, moreover, there is neither disclosure of new vulnerabilities or user data, nor probing of live web services. However, the absence of an explicit ethics-screening step in our methodology means that our moral duty to respect the rights and interests of individuals represented in underlying studies remains only partially fulfilled. We see this as an area of improvement for our future research. At the same time, we upheld other duties by relying exclusively on publicly available artifacts, attributing all sources, avoiding the disclosure of new vulnerabilities, and striving for impartiality in our selection and treatment of studies. Finally, to the best of our knowledge, we complied with regulations, copyrights, and applicable laws. 

\noindent\textbf{Decision to Proceed}
After considering both consequentialist and deontological perspectives, we conclude that conducting and publishing this study constitutes an ethically justified decision. We expect the benefits of this work to outweigh the potential risks and to serve not only the academic community but also the original authors of the reviewed studies and the public at large. Our primary motivation has been to advance knowledge in the domain of web tracker detection and web privacy, and we remain committed to pursuing this goal in the most responsible and transparent manner possible. To foster trust, independent verification, and practical reuse, we release all of our artifacts openly. Finally, we acknowledge the residual risks identified in this analysis and intend to address them in future work by incorporating explicit ethics eligibility criteria into the process of literature inclusion.

\section*{LLM usage considerations}
The work in this study was supported by LLMs for the following use-cases: (i) revising the text; (ii) working with LaTeX; (iii) fixing code issues; (iv) writing a simple web crawler to apply the search matrix to PETS. For (i): LLMs were used for editorial purposes in this manuscript, and all outputs were inspected by the authors to ensure accuracy and originality. 
For (ii): creating table templates and debugging issues (inc. tables) in LaTeX. 
For (iii): the authors encountered several issues during the reproducibility experiments and used LLMs for debugging (in particular dependency issues). For (iv): The PETS website does not offer a search interface to apply search terms. Therefore, we wrote two web crawlers with the help of an LLM to automate this task. Regarding sustainability, each author consciously limited the use of LLMs and shortened queries to minimize the number of tokens. 


\bibliographystyle{IEEEtranS}
\bibliography{IEEEabrv,bibliography}

\appendices
\section{Open Science}
We publicly release all artifacts that were developed during the different phases of this SoK paper, namely: (i) bibliographical data downloaded from digital libraries and academic venues, (ii) source code of the web crawlers, data processing and analysis, (iii) excel sheets with the reviewers decisions for each study from the screening phase, and (iv) code to conduct the limited reproducibility experiments (inc. links to the original repositories)~\cite{zenodo-artifact}. 

\section{Further Information on Systematic Review}\label{app:sys-review}
\subsection{Search Term Matrices and Settings}
We report one example search term matrix for the ACM Digital Library and briefly describe particularities with each source. All search terms are reported in our artifact. 

\noindent\textbf{ACM Digital Library} The following search resulted in 498 papers that were then filtered for research articles (n=105) and short papers (n=7). The search matrix is as follows: 
\begin{tcolorbox}[
  colback=gray!10,        
  coltitle=black,
  frame hidden, 
  sharp corners,
  enhanced,
  left=2mm,               
  boxrule=0pt,         
  borderline west={1pt}{0pt}{snet-red}
]
\small
\textbf{\textsc{Search Matrix (ACM Digital Library):}} [Publication Title: web track* OR "ad blocker" OR first*party* track* OR third*party* track*] OR [Abstract: "web tracker" OR "web tracking" OR "ad tracker" OR "ad blocking" OR first*party* track* OR third*party* track* OR "browser fingerprinting" OR "web tracker detection"] OR [[[Keywords: "web tracker detection" OR "web tracker" OR "web tracking" OR "web privacy" OR "web privacy measurement"]] AND [[NOT [Publication Title: eye* OR blockchain]]]
\end{tcolorbox}

\noindent\textbf{IEEE Xplore} This search resulted in 303 papers, with conference papers (n=283) and journal articles (n=20). The IEEE Xplore search interface did not allow us to use more than ten wildcard characters in one search, which is why we had to use it selectively.   

\noindent\textbf{Scopus} Compared to other digital libraries, the search matrix contains the applied filters. We set the subject area to computer science to reduce the number of results, resulting in conference papers (n=160) and articles (n=57). 

\noindent\textbf{Springer Link} The Springer Link interface did not offer the same possibilities as other digital libraries. We had to enter each word individually in the same order as in the search matrix. The exported file did not contain all relevant information, which required manual effort to add the missing information. Only conference papers (n=98) were exported. 

\noindent\textbf{USENIX} This conference offers their own search interface, albeit limited in its functionality. Results were manually exported and missing information such as URLs and DOIs added. Note that published papers do not have keywords. 

\noindent\textbf{PETS} There was no search interface for PETS, which is why we wrote two crawlers to extract all papers from the proceedings webpage. One for old volumes and one for the newest volume of 2025 as they had a different format than before (November 5\textsuperscript{th}). In addition, volumes before 2015 were not addressed by the crawler as older proceedings were published by Springer International, thus available through the Springer Link digital library. 

\noindent\textbf{arXiv} The arXiv search interface did not allow us to define any search terms for the keywords section. We filtered for studies that were published between 2024-01-01 until 2024-12-31. An export function was not available, so the entire page was saved and results were manually added (incl. missing information) to the review template on November 8\textsuperscript{th}. During our work on this SoK, all identified and relevant arXiv studies have been published. 

\section{Quality of Evidence}\label{sec:quality-results}
To assess the quality of evidence we derive a set of criteria from our systematization (\autoref{sec:performance-evaluation}) that do not require any interpretation from the reviewers (Table~\ref{tab:quality-assess}). These critera can also be used in future studies to ensure a minimum-viable assessment of a detector's performance. The results are as follows: 

\begin{table}[htbp]
  \caption{Evaluation Quality Assessment Questions}
  \label{tab:quality-assess}
  \centering
  \scriptsize
  \begin{tabularx}{\columnwidth}{rX}
    \toprule
    \textbf{Identifier}    & \textbf{Question/Criteria} \\
    \midrule
     & \textbf{Robustness of Metrics} \\
    \midrule
    \textbf{QA1}          & Were Threshold-dependent performance metrics used? \\
    \textbf{QA2}          & Were Threshold-independent performance metrics used?
    \vspace{0.3cm}
    \\
    & \textbf{Baselines} \\
    \midrule
    \textbf{QA3}          & Was the detector compared against a baseline? \\
    \textbf{QA4}          & Was the proposed detector compared an existing detector? 
    \vspace{0.3cm}
    \\
    & \textbf{Robustness} \\
    \midrule
    \textbf{QA5}          & Were possible evasion or circumvention mechanisms discussed or tested? \\
    \textbf{QA6}          & Was a threat model defined?  
    \vspace{0.3cm}
    \\
    & \textbf{Generalizability} \\
    \midrule
    \textbf{QA7}          & Were cross-browser experiments conducted? \\
    \textbf{QA8}          & Were longitudinal experiments conducted? \\
    \textbf{QA9}         & Was the detector only tested in a controlled lab environment? 
    \vspace{0.3cm}
    \\
    & \textbf{Deployment} \\
    \midrule
    \textbf{QA10}         & Was the detector's impact on the quality of experience during browsing considered (inc. breakage and deployment/operations metrics)? \\ 
    \bottomrule
  \end{tabularx}
\end{table}

\noindent\textbf{QA1.} All studies use at least one threshold-dependent performance metric.

\noindent\textbf{QA2.} Only a minority of studies use at least one threshold-independent performance metric (n=7). 

\noindent\textbf{QA3.} Most studies use at least one baseline (n=29). 

\noindent\textbf{QA4.} Some studies use at least one previous detector as a baseline (n=14). 

\noindent\textbf{QA5.} Most studies at least discuss evasion and/or circumvention (n=30). 

\noindent\textbf{QA6.} Only a handful of studies define a threat model (n=5).

\noindent\textbf{QA7.} Only a handful of studies conducted cross-browser experiments (n=3). 

\noindent\textbf{QA8.} Only a minority of studies conducted a longitudinal experiment (n=8). 

\noindent\textbf{QA9.} Most studies conduct lab-only experiments (n=52). 

\noindent\textbf{QA10.} Some studies considered the quality of experience of their detectors. The number is higher than \textbf{QA9} as these considerations can be made without experiments in a real-world environment (n=14). 

\section{Further Information on Reproducibility Experiments}
We reproduce the findings using a server with an AMD EPYC 7713 (64-cores, 128-threads), 256GB RAM, two RTX A6000 with 96GB VRAM in total, 1.8TB SSD and Ubuntu 24 or with a MacBook Pro (14-inch, 2021) with a 10-core CPU, a 16-core GPU, 1TB of SSD storage, and 32GB of RAM. 

\section{Further Information on Systematization}\label{app:labels}

\begin{table}[htbp]
  \caption{Supplementary Studies}
  \label{tab:supplementary-studies}
  \centering
  \scriptsize
  \begin{tabular}{rll}
    \toprule
    \textbf{Year}    & \textbf{Ref} & \textbf{Venue} \\
    \midrule
    1999          & Kushmerick~\cite{kushmerick} & AGENTS\\
    \midrule
    2003          & Alsaid and Martin~\cite{alsaidDetectingWebBugs2003} & PET\\
    \midrule
    2005         & Fonseca et al.~\cite{fonseca} & LA-WEB\\
    \midrule
    2012         & Orr et al.~\cite{orr} & WPES\\
    \midrule
    2014          & Bhagavatula et al.~\cite{bhagavatula} & AISec\\
    \midrule
    2017         & Haga et al.~\cite{hagaBuildingScalableWeb2017} & IEICE Trans. Inf. \& Syst.\\
                 & Yu et al.~\cite{yuEffectivelyProtectYour2017} & CANDAR\\ 
    \midrule
    2018         & Al-Fannah et al.~\cite{al-fannah} & ISC\\ 
                 & Shuba et al.~\cite{shuba} & PETS\\ 
    \midrule
    2021         & Amjad et al.~\cite{amjad-2024} & IMC\\ 
                 & Sjösten et al.~\cite{sjostenEssentialFPExposingEssence2021} & EuroS\&PW\\ 
                 & Zhou and Zhai~\cite{zouBrowserFingerprintingIdentification2021} & HPCC\\ 
    \midrule
    2022         & Li et al.~\cite{fpflow-2022} & CNCERT\\ 
    \midrule
    2023         & Yan et al.~\cite{yanAdhereAutomatedDetection2023} & ICSE\\ 
    \midrule
    2024         & Boussaha et al.~\cite{tracer} & PETS\\ 
                 & Fouad et al.~\cite{fouad-server} & PETS\\ 
                 & Moti et al.~\cite{moti} & IEEE S\&P\\ 
    \bottomrule
  \end{tabular}
\end{table}

\begin{table*}[t]
  \scriptsize
  \renewcommand{\arraystretch}{1.4}
  \caption{Summary of Labeled Statements From the Systematization}
  \begin{tabularx}{\textwidth}{@{}>{\raggedright\arraybackslash}p{10mm}X@{}}
    \toprule
    \textbf{T\#} & \textbf{Trends} \\
    \midrule
    \trend{\textbf{T1}} & Inclusion of threat modeling to assess a detector's robustness or the studied tracking technique and to support derived conclusions. \\
    \trend{\textbf{T2}} & The popularity-based top sites ranking list \textsc{Tranco}~\cite{tranco} is used to identify a suitable sample. \\
    \trend{\textbf{T3}} & \textsc{Selenium} and \textsc{OpenWPM} with different configurations to automate the crawling of  desktop websites with \textsc{Firefox}. \\
    \trend{\textbf{T4}} & Filter lists are the standard for ground truths albeit their known limitations and moderate performance. \\ 
    \trend{\textbf{T5}} & Labeling fingerprinting uses a mix of different methods involving manual input, as filter lists alone are not sufficient. \\
    \trend{\textbf{T6}} & There are real-time and offline modes of detection. \\
    \trend{\textbf{T7}} & Detection is formulated as a supervised binary classification problem, paired with the use of traditional ML-models and curated features. \\
    \trend{\textbf{T8}} & Detectors employ graph-based approaches that target the structural properties of trackers. \\
    \trend{\textbf{T9}} & More experiments and analyses are conducted with a recent focus on generalizability, robustness, and breakage. \\
    \trend{\textbf{T10}} & Detectors are tested in client-side deployment scenarios. \\
    \midrule
    \textbf{L\#} & \textbf{Limitations} \\[4pt]
    \midrule
    \limitation{\textbf{L1}} & Many different crawlers and crawler versions affect reproducibility and accurate measurements. \\
    \limitation{\textbf{L2}} & Filter lists suffer from human errors, biases, noise, and favor functionality over truth for mixed WebTs. This can lead to learned patterns where legitimate behavior is labeled as tracking. \\
    \limitation{\textbf{L3}} & The definition of trackers based on their presence in a filter list results in a coarse-grained circular definition of tracking with limited explanatory power and a biased evaluation towards knowledge already encoded in the lists -- a model is rewarded for re-discovering what is already known. \\
    \limitation{\textbf{L4}} & Detectors may inherit systemic biases from filter lists and risk learning spurious correlates instead of causal tracking behavior. \\
     \limitation{\textbf{L5}} & Detectors are primarily assessed using threshold-dependent performance metrics (PMs) in offline lab evaluations with static data from a single point in time. \\
      \limitation{\textbf{L6}} & The lack of baselines, inappropriate PMs, real-world testing, and diverse evaluation data can misrepresent the performance of detectors and hinder comparability. \\
       \limitation{\textbf{L7}} & Most studies do not evaluate their detectors in real-world scenarios and provide no empirical evidence that blocking decisions preserve site functionality. \\
    \midrule
    \textbf{R\#} & \textbf{Recommendations} \\[4pt]
    \midrule
    \recommendation{\textbf{R1}} & Definition of threat models to clarify a detector's scope and assumptions, and to support robustness evaluations. \\
    \recommendation{\textbf{R2}} & The reporting and crawling are insufficient and should be more carefully planned in the future. \\
    \recommendation{\textbf{R3}} & Future studies should understand, reflect, and mitigate biases of popularity-based ranking lists. \\
    \recommendation{\textbf{R4}} & The long-tail of less-visited websites could be mitigated by a stratified random sampling approach of different ranges such as 1K-10K and 10K-100K . \\
    \recommendation{\textbf{R5}} & Developed heuristics and models should be used to complement common sources for ground truth. \\
    \recommendation{\textbf{R6}} & The labeling definition should reflect the tracker (tracking) definition by the authors. \\
    \recommendation{\textbf{R7}} & Threshold-independent PMs summarize a detector’s overall efficacy and provide a more realistic assessment \\
    \recommendation{\textbf{R8}} & The operational mode, deployment and application scenario should be considered early on in the research process. \\
    \recommendation{\textbf{R9}} & \textsc{Duumviri}'s, \textsc{PURL}'s, and \textsc{AdFlush}'s artifact were reproduced without any issues and are a good example for future studies. \\
    \midrule
    \textbf{O\#} & \textbf{Future Work Opportunities} \\[4pt]
    \midrule
    \opportunity{\textbf{O1}} & The scarceness of threat models leaves open questions about potential vulnerabilities; or evasion and circumvention. \\ 
    \opportunity{\textbf{O2}} & The extent of an adversary's evasion and circumvention; or model manipulations in the real-world are (empirically) unknown. \\ 
    \opportunity{\textbf{O3}} & Popularity-based ranking lists may not result in a representative dataset on web tracking. \\ 
    \opportunity{\textbf{O4}} & A community-standard benchmark suite that combines shared datasets, metrics; and a reference crawling testbed to facilitate comparability, reproducibility, and external validity do not exist yet. \\ 
    \opportunity{\textbf{O5}} & Currently, studies do not benefit from iterative corrections of ground truths made by past studies. \\ 
    \opportunity{\textbf{O6}} & A high-quality ground truth; or unified labeling definition does not exist yet.\\ 
    \opportunity{\textbf{O7}} & Binary labels limit preciseness, explainability, and transparency. \\ 
    \opportunity{\textbf{O8}} & A large body of features has been developed, yet their subsequent use in research remains rare. \\ 
    \opportunity{\textbf{O9}} & Explainable cross-domain and mixture-of-experts detectors with multi-label outputs in a continual learning setting. \\ 
    \opportunity{\textbf{O10}} & Fundamentally more robust, efficient, and user-centric detectors. \\ 
    \opportunity{\textbf{O11}} & The community would benefit from a standardized evaluation framework; an adversarial simulation testbed; and a public benchmark dataset to enable reproducible and valid comparisons between detectors. \\ 
    \opportunity{\textbf{O12}} & Deployment allows for more realistic assessments and opens new, currently under-researched ares in detection. \\ 
    \opportunity{\textbf{O13}} & A comprehensive reproducibility or replicability study with publicly available (containerized) artifacts should be conducted in future work. \\ 
    \bottomrule
  \end{tabularx}
  \label{table:summary-labels}
  \begin{minipage}{\textwidth}
        \vspace{1mm}
        \centering
        \scriptsize
        Each statement from the systematization was summarized and can be found in \autoref{sec:sys}. This summary table was inspired by Wei et al.~\cite{solk} as well. 
        \end{minipage}
\end{table*}

{
\setlength{\tabcolsep}{2.5pt}
\renewcommand{\arraystretch}{0.85}
\begin{table*}[htbp]
    \centering
    \caption{Systematization of Web Tracker Detection Studies and Detectors.} 
    \label{tab:tracker-detector-table}
    \scriptsize
    \begin{adjustbox}{max width=\textwidth}
    \begin{tabular*}{\linewidth}{@{\extracolsep{\fill}} l p{2.6cm} p{1.6cm} *{16}{c} @{}}
            \toprule 
            \multicolumn{3}{c}{\textbf{Bibliographic Info}}
             & \multicolumn{5}{c}{\textbf{Web Artifact}}
             & \multicolumn{6}{c}{\textbf{Feature Type}}
             & \multicolumn{2}{c}{\textbf{Artifact}}
             & \multicolumn{2}{c}{\textbf{Robustness}}
             & \multicolumn{1}{c}{\textbf{Deployment}}
             \\
             \cmidrule(l{-0.05em}r{-0.05em}){1-3}\cmidrule(lr){4-8}\cmidrule(lr){9-14}\cmidrule(lr){15-16}\cmidrule(lr){17-18}\cmidrule(lr){19-19}
             
             \textbf{Year} 
             & \textbf{Ref} 
             & \textbf{Venue}
             & \multicolumn{1}{c}{\textbf{RP}}
             & \multicolumn{1}{c}{\textbf{PM}}
             & \multicolumn{1}{c}{\textbf{WP}}
             & \multicolumn{1}{c}{\textbf{RSC}}
             & \multicolumn{1}{c}{\textbf{SC}}
             & \multicolumn{1}{c}{\textbf{CNT}}
             & \multicolumn{1}{c}{\textbf{STC}}
             & \multicolumn{1}{c}{\textbf{STRCT}}
             & \multicolumn{1}{c}{\textbf{FL}}
             & \multicolumn{1}{c}{\textbf{SEQ}}
             & \multicolumn{1}{c}{\textbf{REL}}
             & \multicolumn{1}{c}{\textbf{Code}}
             & \multicolumn{1}{c}{\textbf{Dataset}}
             & \multicolumn{1}{c}{\textbf{EV-CV}}
             & \multicolumn{1}{c}{\textbf{Threat Model}}
             & \multicolumn{1}{c}{\textbf{}} \\
            \midrule
            2010 
            & Yamada et al.~\cite{yamada} 
            & AINA
            & \ding{55} & \ding{51} & \ding{55} & \ding{55} & \ding{55}
            & \ding{55} & \ding{51} & \ding{51} & \ding{55} & \ding{51} & \ding{51}
             & \ding{55}
             & \ding{55}
             & \ding{55}
             & \ding{55}
             & \ding{55}
            
            \\
            \cmidrule(l{-0.05em}r{-0.05em}){1-3}\cmidrule(lr){4-8}\cmidrule(lr){9-14}\cmidrule(lr){15-16}\cmidrule(lr){17-18}\cmidrule(lr){19-19}
            2013 
            & Bau et al.~\cite{bau} 
            & W2SP
            & \ding{55}
            & \ding{55}
            & \ding{51}
            & \ding{51}
            & \ding{55}
            & \ding{55} & \ding{51} & \ding{51} & \ding{55} & \ding{55} & \ding{51}
             & \ding{55}
             & \ding{55}
             & \ding{51}
             & \ding{55}
             & \ding{55}
            
            \\
            \cmidrule(l{-0.05em}r{-0.05em}){1-3}\cmidrule(lr){4-8}\cmidrule(lr){9-14}\cmidrule(lr){15-16}\cmidrule(lr){17-18}\cmidrule(lr){19-19}
            2015 
            & FaizKhademi et al.~\cite{fpguard} 
            & DBSec
            & \ding{55} & \ding{51} & \ding{51} & \ding{51} & \ding{55}
            & \ding{51} & \ding{51} & \ding{51} & \ding{51} & \ding{55} & \ding{55}
             & \ding{55}
             & \ding{55}
             & \ding{55}
             & \ding{55}
             & \ding{51}
            
            \\
            \addlinespace[1pt]  
            & Gugelmann et al.~\cite{gugelmann} 
            & PETS
            & \ding{55} & \ding{51} & \ding{55} & \ding{55} & \ding{51}
            & \ding{55} & \ding{51} & \ding{55} & \ding{55} & \ding{55} & \ding{51}
             & \ding{55}
             & \ding{55}
             & \ding{55}
             & \ding{55}
             & \ding{55}
            
            \\
            \addlinespace[1pt]  
            & Li et al.~\cite{trackadvisor} 
            & PAM
            & \ding{55} & \ding{51} & \ding{55} & \ding{55} & \ding{51}
            & \ding{51} & \ding{51} & \ding{55} & \ding{55} & \ding{55} & \ding{51}
             & \ding{55}
             & \ding{55}
             & \ding{51}
             & \ding{55}
             & \ding{55}
            
            \\
            \addlinespace[1pt]  
            & Wu et al.~\cite{trackerdetector} 
            & Comp. Netw.
            & \ding{55} & \ding{55} & \ding{51} & \ding{51} & \ding{51}
            & \ding{51} & \ding{51} & \ding{55} & \ding{51} & \ding{55} & \ding{55}
             & \ding{55}
             & \ding{55}
             & \ding{51}
             & \ding{55}
             & \ding{55}
            
            \\
            \addlinespace[1pt]  
            & Metwalley et al.~\cite{metwalley} 
            & GLOBECOM
            & \ding{51} & \ding{51} & \ding{55} & \ding{55} & \ding{51}
            & \ding{51} & \ding{51} & \ding{55} & \ding{51} & \ding{55} & \ding{51}
             & \ding{55}
             & \ding{55}
             & \ding{55}
             & \ding{55}
             & \ding{55}
            
            \\
            \cmidrule(l{-0.05em}r{-0.05em}){1-3}\cmidrule(lr){4-8}\cmidrule(lr){9-14}\cmidrule(lr){15-16}\cmidrule(lr){17-18}\cmidrule(lr){19-19}
            2016 
            & Wu et al.~\cite{wu2016} 
            & ESORICS
            & \ding{51} & \ding{55} & \ding{51} & \ding{55} & \ding{51}
            & \ding{51} & \ding{51} & \ding{51} & \ding{51} & \ding{55} & \ding{51}
             & \ding{55}
             & \ding{55}
             & \ding{55}
             & \ding{55}
             & \ding{55}
            
            \\
            \addlinespace[1pt]  
            & Dudykevych and Nechypor~\cite{dudykevych} 
            & PIC S\&T
            & \ding{55} & \ding{51} & \ding{55} & \ding{55} & \ding{51}
            & \ding{51} & \ding{51} & \ding{55} & \ding{51} & \ding{55} & \ding{51}
             & \ding{55}
             & \ding{55}
             & \ding{55}
             & \ding{55}
             & \ding{55}
            
            \\
            \addlinespace[1pt]  
            & Kalavri et al.~\cite{kalavri} 
            & PAM
            & \ding{51} & \ding{51} & \ding{55} & \ding{51} & \ding{55}
            & \ding{55} & \ding{51} & \ding{51} & \ding{55} & \ding{55} & \ding{51}
             & \ding{55}
             & \ding{55}
             & \ding{55}
             & \ding{55}
             & \ding{55}
            
            \\
            \addlinespace[1pt]  
            & Kaizer and Gupta.~\cite{kaizer} 
            & IWSPA
            & \ding{51} & \ding{51} & \ding{51} & \ding{55} & \ding{51}
            & \ding{55} & \ding{51} & \ding{55} & \ding{55} & \ding{55} & \ding{51} 
             & \ding{55}
             & \ding{55}
             & \ding{55}
             & \ding{55}
             & \ding{55}
            
            \\
            \addlinespace[1pt]  
            & Yu et al.~\cite{yu} 
            & WWW
            & \ding{51} & \ding{55} & \ding{51} & \ding{55} & \ding{51}
            & \ding{55} & \ding{51} & \ding{55} & \ding{55} & \ding{55} & \ding{51}
             & \ding{55}
             & \ding{55}
             & \ding{51}
             & \ding{55}
             & \ding{51}
            
            \\
            \cmidrule(l{-0.05em}r{-0.05em}){1-3}\cmidrule(lr){4-8}\cmidrule(lr){9-14}\cmidrule(lr){15-16}\cmidrule(lr){17-18}\cmidrule(lr){19-19}
            2017 
            & Ikram et al.~\cite{ikram} 
            & PETS
            & \ding{55} & \ding{55} & \ding{51} & \ding{51} & \ding{51}
            & \ding{51} & \ding{51} & \ding{51} & \ding{55} & \ding{51} & \ding{55} 
             & \ding{55}
             & \ding{55}
             & \ding{51}
             & \ding{55}
             & \ding{55}
            
            \\
            \cmidrule(l{-0.05em}r{-0.05em}){1-3}\cmidrule(lr){4-8}\cmidrule(lr){9-14}\cmidrule(lr){15-16}\cmidrule(lr){17-18}\cmidrule(lr){19-19}
            2018 
            & Sanchez-Rola and Santos~\cite{sanchez2018} 
            & DIMVA
            & \ding{55} & \ding{55} & \ding{51} & \ding{51} & \ding{55}
            & \ding{51} & \ding{51} & \ding{55} & \ding{55} & \ding{55} & \ding{55}
             & \ding{55}
             & \ding{55}
             & \ding{51}
             & \ding{55}
             & \ding{55}
            
            \\
            \cmidrule(l{-0.05em}r{-0.05em}){1-3}\cmidrule(lr){4-8}\cmidrule(lr){9-14}\cmidrule(lr){15-16}\cmidrule(lr){17-18}\cmidrule(lr){19-19}
            2019 
            & Zhao et al.~\cite{zhao} 
            & SecureComm
            & \ding{51} & \ding{51} & \ding{55} & \ding{55} & \ding{51}
            & \ding{55} & \ding{51} & \ding{51} & \ding{51} & \ding{55} & \ding{55}
             & \ding{55}
             & \ding{55}
             & \ding{55}
             & \ding{55}
             & \ding{55}
            
            \\
            \addlinespace[1pt]  
            & Vo and Jaiswal~\cite{adremover} 
            & UEMCON
            & \ding{51} & \ding{55} & \ding{55} & \ding{55} & \ding{55}
            & \ding{51} & \ding{51} & \ding{55} & \ding{51} & \ding{55} & \ding{51}
             & \ding{55}
             & \ding{55}
             & \ding{55}
             & \ding{55}
             & \ding{55}
            
            \\
            \cmidrule(l{-0.05em}r{-0.05em}){1-3}\cmidrule(lr){4-8}\cmidrule(lr){9-14}\cmidrule(lr){15-16}\cmidrule(lr){17-18}\cmidrule(lr){19-19}
            2020 
            & Dao and Fukada~\cite{dao} 
            & GLOBECOM
            & \ding{51} & \ding{51} & \ding{51} & \ding{51} & \ding{55}
            & \ding{51} & \ding{51} & \ding{51} & \ding{51} & \ding{55} & \ding{51}
             & \ding{55}
             & \ding{51}
             & \ding{51}
             & \ding{55}
             & \ding{55}
            
            \\
            \addlinespace[1pt]  
            & Castell-Uroz et al.~\cite{uroz-dl} 
            & CNSM
            & \ding{51} & \ding{55} & \ding{55} & \ding{55} & \ding{55}
            & \ding{51} & \ding{55} & \ding{55} & \ding{55} & \ding{51} & \ding{55}
             & \ding{55}
             & \ding{55}
             & \ding{55}
             & \ding{55}
             & \ding{55}
            
            \\
            \addlinespace[1pt]  
            & Iqbal et al.~\cite{adgraph} 
            & S\&P
            & \ding{51} & \ding{51} & \ding{51} & \ding{51} & \ding{55}
            & \ding{51} & \ding{51} & \ding{51} & \ding{51} & \ding{55} & \ding{51}
             & \ding{51}
             & \ding{55}
             & \ding{51}
             & \ding{55}
             & \ding{55}
            
            \\
            \addlinespace[1pt]  
            & Cozza et al.~\cite{cozza} 
            & Comp. Netw.
            & \ding{51} & \ding{51} & \ding{51} & \ding{55} & \ding{51}
            & \ding{51} & \ding{51} & \ding{55} & \ding{51} & \ding{55} & \ding{51}
             & \ding{55}
             & \ding{55}
             & \ding{51}
             & \ding{55}
             & \ding{55}
            
            \\
            \addlinespace[1pt]  
            & Guarino et al.~\cite{guarino} 
            & ICISSP
            & \ding{51} & \ding{51} & \ding{55} & \ding{55} & \ding{51}
            & \ding{51} & \ding{51} & \ding{55} & \ding{55} & \ding{55} & \ding{51} 
             & \ding{55}
             & \ding{55}
             & \ding{55}
             & \ding{55}
             & \ding{55}
            
            \\
            \cmidrule(l{-0.05em}r{-0.05em}){1-3}\cmidrule(lr){4-8}\cmidrule(lr){9-14}\cmidrule(lr){15-16}\cmidrule(lr){17-18}\cmidrule(lr){19-19}
            2021 
            & Sun et al.~\cite{sun-2021} 
            & INFOCOM
            & \ding{51} & \ding{55} & \ding{51} & \ding{51} & \ding{51}
            & \ding{51} & \ding{51} & \ding{55} & \ding{51} & \ding{55} & \ding{51}
             & \ding{55}
             & \ding{55}
             & \ding{55}
             & \ding{55}
             & \ding{55}
            
            \\
            \addlinespace[1pt]  
            & Hu et al.~\cite{cccc-2021} 
            & WebSci
            & \ding{55} & \ding{55} & \ding{55} & \ding{55} & \ding{51}
            & \ding{51} & \ding{55} & \ding{55} & \ding{55} & \ding{51} & \ding{55}
             & \ding{55}
             & \ding{55}
             & \ding{51}
             & \ding{55}
             & \ding{55}
            
            \\
            \addlinespace[1pt]  
            & Chen et al.~\cite{chen-2021} 
            & S\&P
            & \ding{51} & \ding{55} & \ding{51} & \ding{51} & \ding{51}
            & \ding{55} & \ding{55} & \ding{51} & \ding{51} & \ding{51} & \ding{55}
             & \ding{51}
             & \ding{51}
             & \ding{51}
             & \ding{55}
             & \ding{55}
            
            \\
            \addlinespace[1pt]  
            & Kargaran et al.~\cite{kargaran-2021} 
            & WebSci
            & \ding{51} & \ding{55} & \ding{55} & \ding{51} & \ding{55}
            & \ding{51} & \ding{51} & \ding{51} & \ding{51} & \ding{55} & \ding{51} 
             & \ding{55}
             & \ding{55}
             & \ding{51}
             & \ding{55}
             & \ding{55}
            
            \\
            \addlinespace[1pt]  
            & Reitinger and Mazurek~\cite{reitinger-ml-cb_2021} 
            & PETS
            & \ding{55} & \ding{55} & \ding{51} & \ding{51} & \ding{55}
            & \ding{51} & \ding{51} & \ding{55} & \ding{55} & \ding{51} & \ding{55}
             & \ding{55}
             & \ding{55}
             & \ding{51}
             & \ding{55}
             & \ding{55}
            
            \\
            \addlinespace[1pt]  
            & Rizzo et al.~\cite{rizzo-unveiling-2021} 
            & PETS
            & \ding{55} & \ding{55} & \ding{51} & \ding{51} & \ding{55}
            & \ding{51} & \ding{51} & \ding{51} & \ding{55} & \ding{55} & \ding{55}
             & \ding{55}
             & \ding{55}
             & \ding{51}
             & \ding{55}
             & \ding{55}
            
            \\
            \addlinespace[1pt]  
            & Iqbal et al.~\cite{iqbal-fingerprinting-2021} 
            & S\&P
            & \ding{55} & \ding{51} & \ding{51} & \ding{51} & \ding{55}
            & \ding{51} & \ding{51} & \ding{55} & \ding{51} & \ding{51} & \ding{51}
             & \ding{51}
             & \ding{55}
             & \ding{55}
             & \ding{55}
             & \ding{55}
            
            \\
            \addlinespace[1pt]  
            & Castell-Uroz et al.~\cite{castell-uroz-tracksign_2021} 
            & INFOCOM
            & \ding{55} & \ding{55} & \ding{51} & \ding{51} & \ding{55}
            & \ding{51} & \ding{51} & \ding{51} & \ding{55} & \ding{55} & \ding{51}
             & \ding{51}
             & \ding{51}
             & \ding{55}
             & \ding{55}
             & \ding{55}
            
            \\
            \addlinespace[1pt]  
            & Dao et al.\textsuperscript{\textdagger}~\cite{dao-cname-journal} 
            & IEEE TNSM
            & \ding{51} & \ding{51} & \ding{51} & \ding{51} & \ding{55}
            & \ding{51} & \ding{51} & \ding{55} & \ding{55} & \ding{55} & \ding{51}
             & \ding{55}
             & \ding{51}
             & \ding{51}
             & \ding{55}
             & \ding{51}
            
            \\
            \cmidrule(l{-0.05em}r{-0.05em}){1-3}\cmidrule(lr){4-8}\cmidrule(lr){9-14}\cmidrule(lr){15-16}\cmidrule(lr){17-18}\cmidrule(lr){19-19}
            2022 
            & Yang et al.~\cite{yang_wtagraph_2022} 
            & S\&P
            & \ding{51} & \ding{51} & \ding{51} & \ding{55} & \ding{51}
            & \ding{51} & \ding{51} & \ding{51} & \ding{55} & \ding{51} & \ding{51}
             & \ding{51}
             & \ding{51}
             & \ding{51}
             & \ding{55}
             & \ding{51}
            
            \\
            \addlinespace[1pt]  
            & Cheng~\cite{cheng_using_2022} 
            & ICCSMT
            & \ding{55} & \ding{55} & \ding{51} & \ding{55} & \ding{55}
            & \ding{51} & \ding{51} & \ding{55} & \ding{55} & \ding{51} & \ding{55}
             & \ding{55}
             & \ding{55}
             & \ding{55}
             & \ding{55}
             & \ding{55}
            
            \\
            \addlinespace[1pt]  
            & M et al.\cite{krishna-2022} 
            & ICCCNT
            & \ding{51} & \ding{51} & \ding{51} & \ding{55} & \ding{55}
            & \ding{51} & \ding{51} & \ding{55} & \ding{55} & \ding{55} & \ding{55} 
             & \ding{51}
             & \ding{55}
             & \ding{51}
             & \ding{55}
             & \ding{55}
            
            \\
            \addlinespace[1pt]  
            & Siby et al.~\cite{siby-2022} 
            & USENIX
            & \ding{55} & \ding{51} & \ding{51} & \ding{55} & \ding{51}
            & \ding{55} & \ding{51} & \ding{51} & \ding{51} & \ding{55} & \ding{51}
             & \ding{51}
             & \ding{55}
             & \ding{51}
             & \ding{51}
             & \ding{55}
            
            \\
            \addlinespace[1pt]  
            & Iqbal et al.~\cite{khaleesi-2022} 
            & USENIX 
            & \ding{51} & \ding{51} & \ding{55} & \ding{55} & \ding{51}
            & \ding{51} & \ding{51} & \ding{55} & \ding{51} & \ding{51} & \ding{51}
             & \ding{51}
             & \ding{51}
             & \ding{51}
             & \ding{55}
             & \ding{51}
            
            \\
            \addlinespace[1pt]  
            & Bahrami et al.~\cite{bahrami_fp-radar_2022} 
            & PETS 
            & \ding{55} & \ding{55} & \ding{51} & \ding{51} & \ding{55}
            & \ding{51} & \ding{51} & \ding{51} & \ding{55} & \ding{55} & \ding{55}
             & \ding{55}
             & \ding{55}
             & \ding{51}
             & \ding{55}
             & \ding{55}
            
            \\
            \cmidrule(l{-0.05em}r{-0.05em}){1-3}\cmidrule(lr){4-8}\cmidrule(lr){9-14}\cmidrule(lr){15-16}\cmidrule(lr){17-18}\cmidrule(lr){19-19}
            2023 
            & Lee and Son~\cite{lee_adcpg_2023} 
            & CCS
            & \ding{55} & \ding{55} & \ding{51} & \ding{55} & \ding{51}
            & \ding{51} & \ding{55} & \ding{51} & \ding{51} & \ding{55} & \ding{55}
             & \ding{51}
             & \ding{55}
             & \ding{55}
             & \ding{55}
             & \ding{51}
            
            \\
            \addlinespace[1pt]  
            & Ghasemisharif and Polakis~\cite{bytecode-2023} 
            & CCS
            & \ding{55} & \ding{55} & \ding{55} & \ding{51} & \ding{55}
            & \ding{51} & \ding{51} & \ding{55} & \ding{55} & \ding{51} & \ding{55} 
             & \ding{51}
             & \ding{51}
             & \ding{51}
             & \ding{55}
             & \ding{55}
            
            \\
            \addlinespace[1pt]  
            & Munir et al.~\cite{cookiegraph-2023} 
            & CCS
            & \ding{51} & \ding{51} & \ding{51} & \ding{51} & \ding{51}
            & \ding{55} & \ding{51} & \ding{51} & \ding{51} & \ding{55} & \ding{51}
             & \ding{51}
             & \ding{55}
             & \ding{51}
             & \ding{51}
             & \ding{55}
            
            \\
            \addlinespace[1pt]  
            & Zhao~\cite{zhao_fprobe_2023} 
            & ICCCN
            & \ding{55} & \ding{55} & \ding{51} & \ding{51} & \ding{55}
            & \ding{51} & \ding{51} & \ding{55} & \ding{55} & \ding{55} & \ding{51}
             & \ding{55}
             & \ding{55}
            & \ding{55} 
            & \ding{55}
             & \ding{55}
            
            \\
            \addlinespace[1pt]  
            & Su and Kpravelos~\cite{su_automatic_2023} 
            & WWW
            & \ding{55} & \ding{55} & \ding{51} & \ding{55} & \ding{51}
            & \ding{51} & \ding{55} & \ding{55} & \ding{51} & \ding{51} & \ding{55}
             & \ding{51}
             & \ding{51}
             & \ding{51}
             & \ding{55}
             & \ding{55}
            
            \\
            \addlinespace[1pt]  
            & Lee et al.~\cite{lee_net-track_2023} 
            & WWW
            & \ding{55} & \ding{51} & \ding{55} & \ding{55} & \ding{55}
            & \ding{55} & \ding{51} & \ding{55} & \ding{55} & \ding{51} & \ding{55} 
             & \ding{55}
             & \ding{55}
             & \ding{51}
             & \ding{55}
             & \ding{55}
            
            \\
            \addlinespace[1pt]  
            & Ahmad et al.~\cite{ahmad_empirical_2023} 
            & TWEB
            & \ding{51} & \ding{51} & \ding{51} & \ding{55} & \ding{51}
            & \ding{51} & \ding{55} & \ding{55} & \ding{51} & \ding{55} & \ding{51}
             & \ding{55}
             & \ding{55}
             & \ding{51}
             & \ding{55}
             & \ding{55}
            
            \\
            \addlinespace[1pt]  
            & Polcǎk et al.~\cite{polcak_jshelter_2023} 
            & SECRYPT
            & \ding{55} & \ding{55} & \ding{51} & \ding{55} & \ding{55}
            & \ding{51} & \ding{51} & \ding{55} & \ding{55} & \ding{55} & \ding{55} 
             & \ding{55}
             & \ding{55}
             & \ding{51}
             & \ding{51}
             & \ding{51}
            
            \\
            \addlinespace[1pt]  
            & Raschke et al.~\cite{raschke_tex-graph_2023} 
            & ICISSP
            & \ding{51} & \ding{51} & \ding{55} & \ding{51} & \ding{51}
            & \ding{51} & \ding{51} & \ding{51} & \ding{55} & \ding{55} & \ding{51}
             & \ding{51}
             & \ding{51}
             & \ding{55}
             & \ding{55}
             & \ding{55}
            
            \\
            \addlinespace[1pt]  
            & Castell-Uroz et al.\textsuperscript{\textdagger}~\cite{castell-uroz_early_2023} 
            & Comput. Commun.
            & \ding{55} & \ding{55} & \ding{51} & \ding{51} & \ding{55}
             & \ding{51} & \ding{55} & \ding{51} & \ding{55} & \ding{55} & \ding{51} 
             & \ding{51}
             & \ding{51}
             & \ding{51}
             & \ding{55}
             & \ding{55}
            
            \\
            \addlinespace[1pt]  
            & Castell-Uroz et al.~\cite{castell-uroz_astrack_2023} 
            & INFOCOM
            & \ding{51} & \ding{55} & \ding{51} & \ding{51} & \ding{55}
            & \ding{51} & \ding{51} & \ding{55} & \ding{55} & \ding{55} & \ding{51} 
             & \ding{51}
             & \ding{51}
             & \ding{51}
             & \ding{55}
             & \ding{55}
            
            \\
            \cmidrule(l{-0.05em}r{-0.05em}){1-3}\cmidrule(lr){4-8}\cmidrule(lr){9-14}\cmidrule(lr){15-16}\cmidrule(lr){17-18}\cmidrule(lr){19-19}
            2024 
            & Amjad et al.~\cite{amjad-2024} 
            & CCS 
            & \ding{51} & \ding{51} & \ding{51} & \ding{55} & \ding{51}
            & \ding{55} & \ding{51} & \ding{51} & \ding{51} & \ding{51} & \ding{55}
             & \ding{55}
             & \ding{55}
             & \ding{51}
             & \ding{51}
             & \ding{55}
            
            \\
            \addlinespace[1pt]  
            & Campeny-Roig et al.~\cite{campeny-2024} 
            & GNNET
            & \ding{55} & \ding{55} & \ding{55} & \ding{51} & \ding{55}
            & \ding{51} & \ding{55} & \ding{51} & \ding{55} & \ding{55} & \ding{55}
             & \ding{55}
             & \ding{55}
             & \ding{51}
             & \ding{55}
             & \ding{55}
            
            \\
            \addlinespace[1pt]  
            & Annamalai et al.~\cite{annamalai-2024} 
            & NDSS
            & \ding{55} & \ding{55} & \ding{51} & \ding{55} & \ding{55}
            & \ding{51} & \ding{51} & \ding{55} & \ding{55} & \ding{55} & \ding{55}
             & \ding{55}
             & \ding{55}
             & \ding{51}
             & \ding{55}
             & \ding{51}
            
            \\
            \addlinespace[1pt]  
            & Munir and Lee.~\cite{munir_purl_2024} 
            & USENIX
            & \ding{51} & \ding{55} & \ding{51} & \ding{55} & \ding{51}
            & \ding{51} & \ding{51} & \ding{51} & \ding{51} & \ding{51} & \ding{55} 
             & \ding{51}
             & \ding{51}
             & \ding{51}
             & \ding{51}
             & \ding{55}
            
            \\
            \addlinespace[1pt]  
            & Lee et al.~\cite{lee_adflush_2024} 
            & WWW
            & \ding{51} & \ding{51} & \ding{51} & \ding{55} & \ding{51}
            & \ding{51} & \ding{51} & \ding{51} & \ding{55} & \ding{55} & \ding{51} 
             & \ding{51}
             & \ding{51}
             & \ding{51}
             & \ding{55}
             & \ding{51}
            
            \\
            \addlinespace[1pt]  
            & Qiang et al.~\cite{qiang_deepfpd_2024} 
            & Trans. Reliab.
            & \ding{55} & \ding{55} & \ding{51} & \ding{51} & \ding{55}
            & \ding{51} & \ding{51} & \ding{51} & \ding{55} & \ding{51} & \ding{55}
             & \ding{55}
             & \ding{55}
             & \ding{51}
             & \ding{55}
             & \ding{55}
            
            \\
            \addlinespace[1pt]  
            & Wittig and Kesdoğan~\cite{meyer_detecting_2024} 
            & SEC
            & \ding{55} & \ding{51} & \ding{55} & \ding{55} & \ding{55}
            & \ding{55} & \ding{51} & \ding{55} & \ding{51} & \ding{51} & \ding{51}
             & \ding{51}
             & \ding{51}
             & \ding{51}
             & \ding{55}
             & \ding{55}
            
            \\
            \addlinespace[1pt]  
            & Zimmeck et al.~\cite{zimmeck_website_2024} 
            & PETS 
            & \ding{51} & \ding{51} & \ding{51} & \ding{51} & \ding{51}
            & \ding{51} & \ding{51} & \ding{55} & \ding{51} & \ding{51} & \ding{51}
             & \ding{51}
             & \ding{55}
             & \ding{55}
             & \ding{55}
             & \ding{55}
            
            \\
            \addlinespace[1pt]  
            & Yuan et al.~\cite{ast-trans} 
            & IPCCC 
            & \ding{55} & \ding{55} & \ding{51} & \ding{51} & \ding{55}
            & \ding{51} & \ding{51} & \ding{51} & \ding{55} & \ding{51} & \ding{55}
             & \ding{55}
             & \ding{55}
             & \ding{51}
             & \ding{55}
             & \ding{55}
            
            \\
            \cmidrule(l{-0.05em}r{-0.05em}){1-3}\cmidrule(lr){4-8}\cmidrule(lr){9-14}\cmidrule(lr){15-16}\cmidrule(lr){17-18}\cmidrule(lr){19-19}
            2025 
            & Rieder et al.~\cite{rieder_beyond_2025} 
            & PETS 
            & \ding{55} & \ding{51} & \ding{55} & \ding{55} & \ding{55}
            & \ding{51} & \ding{51} & \ding{55} & \ding{55} & \ding{55} & \ding{51}
             & \ding{51}
             & \ding{51}
             & \ding{51}
             & \ding{55}
             & \ding{55}
            
            \\
            \addlinespace[1pt]  
            & Shuang et al.~\cite{shuang_dumviri_2025} 
            & NDSS 
            & \ding{55} & \ding{51} & \ding{51} & \ding{55} & \ding{51}
            & \ding{51} & \ding{51} & \ding{51} & \ding{51} & \ding{51} & \ding{51}
             & \ding{51}
             & \ding{51}
             & \ding{51}
             & \ding{55}
             & \ding{51}
            
            \\
            \addlinespace[1pt] 
            & Wang et al.~\cite{wang-2025} 
            & ADMA 
            & \ding{55} & \ding{55} & \ding{55} & \ding{51} & \ding{55}
            & \ding{51} & \ding{51} & \ding{51} & \ding{55} & \ding{55} & \ding{55}
             & \ding{55}
             & \ding{55}
             & \ding{55}
             & \ding{55}
             & \ding{55}
            
            \\
            \bottomrule 
        \end{tabular*}
    \end{adjustbox}
    \begin{minipage}{\textwidth}
    \vspace{1mm}
    \centering
    \scriptsize
    The table summarizes the identified detectors along a subset of the most relevant characteristics that are further explained in Section~\ref{sec:sys} and their publication year. \textbf{Header abbreviations:} request path (RP); protocol metadata (PM); web page (WP); resource (RSC); storage \& cache (SC); content (CNT); statistical (STC); structural (STRCT); sequential (SEQ); relation (REL); filter list (FL); evasion and circumvention (EV-CV). \textbf{Symbols:} \textdagger refers to journal extensions of listed papers; \ding{51} included; \ding{55} not included. 
    \end{minipage}
\end{table*}}

\clearpage
\newpage 

\section{Meta-Review}

The following meta-review was prepared by the program committee for the 2026
IEEE Symposium on Security and Privacy (S\&P) as part of the review process as
detailed in the call for papers. 

\subsection{Summary}
The submission is an SoK for web tracker detection research. The submission presents its methodology for identifying relevant papers and introduces a taxonomy based on an ML-style workflow. The systematization includes 59 papers and also lists research opportunities and limitations. Finally, the submission reproduces seven works that present sufficient material for such a reproduction. 

\subsection{Scientific Contributions}
\begin{itemize}
\item Provides a Valuable Step Forward in an Established Field.
\end{itemize}

\subsection{Reasons for Acceptance}
\begin{enumerate}
\item The paper undertakes a comprehensive review of papers on web tracking detection research. The methodology, taxonomy, and systematization are generally detailed and provide a good overview of the ecosystem.
\item The paper reproduces some of the papers and provides some suggestions on artifact issues, which make sense.
\item The ethics considerations are detailed and should be helpful to authors of SoK papers in general.
\end{enumerate}

\end{document}